\documentclass[a4paper,12pt]{article}

\usepackage{amsthm}{\normalsize }
\usepackage{amsmath}
\usepackage{amssymb}
\usepackage{latexsym}
\usepackage{float}
\usepackage{diagbox} 
\usepackage{threeparttable}
\usepackage[textwidth=18cm,textheight=20cm]{geometry}
\usepackage{graphicx} 
\usepackage[square,sort,comma,numbers]{natbib}
\newtheorem{theorem}{Theorem}[section]
\newtheorem{proposition}[theorem]{Proposition}
\newtheorem{lemma}[theorem]{Lemma}
\newtheorem{corollary}[theorem]{Corollary}
\newtheorem{remark}[theorem]{Remark}

\begin{document}


\title{A note on closed-form spread option valuation under log-normal models} 


\author{
Nuerxiati Abudurexiti \thanks{Xi'an Jiaotong Liverpool University, Suzhou, China.\newline  Email: N.Abudurexiti19@student.xjtlu.edu.cn} \and Kai He \thanks{Xi'an Jiaotong Liverpool University, Suzhou, China.\newline Kai.he18@student.xjtlu.edu.cn}\and 
Dongdong Hu \thanks{Xi'an Jiaotong Liverpool University, Suzhou, China.\newline Email: Dongdong.Hu20@student.xjtlu.edu.cn}  \and
Hasanjan Sayit \thanks{Xi'an Jiaotong Liverpool University, Suzhou, China.\newline Email: Hasanjan Sayit@xjtlu.edu.cn}}

\maketitle

\begin{abstract}
In the papers \cite{CarmonaandDurrleman2003a} and \cite{bjerksund2014closed}, closed-form approximations for spread call option prices were studied under the log-normal models. In this paper, we give an alternative closed-form formula for the price of spread call options under the log-normal models also. Our formula can be seen as a generalization of the closed-form formula presented in  \cite{bjerksund2014closed} as their formula can be obtained by selecting special parameter values for our formula. Numerical tests show that our formula performs better for a certain range of model parameters than the closed-form formula presented in \cite{bjerksund2014closed}.
\end{abstract}



\section{Introduction}

A spread option is a contract that gives the owner the right, but not the obligation, to receive the difference, or spread, between the prices of two assets. Spread options can be written on various types of financial products including equities, bonds, currencies, and commodities. Their use is widespread in fixed income,   currency, commodity futures, and energy markets. As such pricing and hedging techniques of spread options are very important for the financial industry. 

One of the difficulties in pricing spread options is that the exercise boundary is not in a linear form when the strike price is not zero. When the strike price of the spread option is zero, a spread option becomes an exchange option for which a closed-form formula exists. This formula is called Margrabe  formula in the literature (see \cite{margrabe1978value}). When the strike price of the spread option is not zero, the exercise boundary is given by a curve and to obtain spread option price one needs to evaluate the risk-neutral probability that the joint log prices lie in one side of that curve, see our Proposition \ref{prop1} below. Due to this non-linearity of the exercise boundary, it is  challenging to compute spread option prices efficiently and accurately.

Numerous papers were devoted in pricing spread options in the past either by deriving approximate analytical closed-form formulas or by  introducing various numerical methods. The most popular numerical method in pricing spread options is probably the application of the fast Fourier Transformation technique (FFT) which was introduced in \citep{carr1999option} initially for pricing European options written on  single asset. The idea of this FFT technique was later extended to the case of pricing problems of spread options; see \cite{dempster2002spread} and \cite{hurd2010fourier} for example. Both of these papers apply FFT in bi-variate Fourier transformation setting. The recent papers \cite{caldana2012spread, caldana2016general, Caldana_Fusai20134893}  studied pricing problems of spread options under general price dynamics by using Fourier Transformation techniques also. In their papers, they applied the FFT technique to a lower bound for the spread option price that were obtained by using the similar idea in the paper  \cite{bjerksund2014closed}. By doing so they were able to give an approximate semi-closed form formula for the spread option prices. Their approach gives spread prices by a uni-variante Fourier inversion formula which is easier and faster to compute than the formulas that involve bi-variate Fourier Transformation in \cite{dempster2002spread} and \cite{hurd2010fourier}.

Analytical methods offer closed-form approximate formulas to spread option prices. The first such approximation is the Bachelier approximation (see \cite{wilcox1990energy} and \cite{poitras1998spread}). In the Bachelier approximation, the price difference of two assets is approximated by a normal random variable. This approximation gives a closed-form formula for the spread option prices but it does not give very accurate spread prices. Another closed-form formula for spread options was proposed by \cite{kirk1995correlation}. This formula has been popular among practitioners as it is simple and gives relatively accurate spread prices. The Kirk's closed-form formula is obtained by approximating the sum of the second asset with the fixed strike by a log-normal random variable. After this approximation is implemented, Kirk's formula was obtained by using a similar approach in calculating \cite{margrabe1978value} formula. The papers \cite{CarmonaandDurrleman2003a, CarmonaandDurrleman2003b} introduced a new approach in approximating spread option prices in closed-form under log-normal models also. They gave lower and upper bounds to the spread prices. The \cite{CarmonaandDurrleman2003a} spread option formula gives very accurate spread option prices, but it involves solving a nonlinear system of equations, which is not trivial.

The recent paper by \cite{bjerksund2014closed} also developed a closed-form formula for spread option prices under log-normal models. They first gave a lower bound to the spread option price and obtained their closed form formula by computing  
this lower bound. Numerical tests show that the formula in \cite{bjerksund2014closed} performs better than the Kirk's formula, see Section 7 of \cite{bjerksund2014closed} for this.

The purpose of this paper is to give an alternative closed-form analytical formula for spread options under the log-normal models also. Our formula can be seen as a generalization of the formula in \cite{bjerksund2014closed}, as their formula can be obtained by plugging in some special parameter values into our closed-form formula. Numerical tests show that our formula performs better than the closed formula in \cite{bjerksund2014closed} for a large range of model parameters.

The paper is organized as follows. In Section 2 we describe our model. In Section 3, we review the \cite{bjerksund2014closed} and \cite{CarmonaandDurrleman2003a} formulas. In Section 4, we present two propositions, Proposition \ref{prop1} and Proposition \ref{prop2}, that are useful for our discussions in the remaining sections of the paper. Especially, our Proposition \ref{prop2} gives a numerical procedure which performs faster compared to the Monte Carlo methods. Section 5 presents the main results of this paper. In this section, we obtain a closed-form formula for spread option price (see Proposition \ref{mainmain}) and show that the closed-form formula for spread options presented in \cite{bjerksund2014closed} is just a special case of our formula. Section 6 presents numerical results, and the Appendix section presents proofs of our results.

\section{The Problem}
We consider a financial market with one risk-free asset with a constant interest rate $r$ and two stocks in a finite time horizon $[0, T]$. The price dynamics of the two stocks under the risk neutral measure are given by
\begin{equation}\label{1}
\begin{split}
dS_1(t)=(r-q_1)S_1(t)dt+\sigma_1S_1(t)dW_1(t),\\
dS_2(t)=(r-q_2)S_2(t)dt+\sigma_2S_2(t)dW_2(t),
\end{split}
\end{equation}
 where $q_1$ and $q_2$ are instantaneous dividend yields of the two stocks respectively, $\sigma_1$ and $\sigma_2$ are positive constants that represent the volatilities of these two stocks, and $W_1(t)$ and $W_2(t)$ are two Brownian motions with correlation $\rho$  under the risk-neutral measure $Q$.
 For the details of risk-neutral measure and risk-neutral pricing see \cite{delbaen1994general, harrison1983stochastic, jarrow2007introduction, black1976pricing} and the references therein.

 We denote by $S_1(0)$ and $S_2(0)$ the stock price at time $t=0$ respectively and we let $F_1=S_1(0)e^{(r-q_1)T}$ and $F_2=S_2(0)e^{(r-q_2)T}$ denote the forward prices. Then the stock prices at time $T$ can be 
 written as
\begin{equation}\label{three}
 S_1(T)=F_1e^{-\frac{1}{2}\sigma^2_1T+\sigma_1W_1(T)}, \;   S_2(T)=F_2e^{-\frac{1}{2}\sigma^2_2T+\sigma_2W_2(T)}. 
\end{equation}
The price of a spread option  with strike price $K$ is given by
\begin{equation}\label{spreadprice}
\Pi_T=\Pi_T(F_1, \sigma_1, F_2, \sigma_2, K, \rho)=e^{-rT}E[(S_1(T)-S_2(T)-K)^+],
\end{equation}
where the expectation is taken under the risk-neutral measure $Q$. In (\ref{spreadprice}) we used a similar notation as in \cite{CarmonaandDurrleman2003a} for the spread price. By using the parity relation discussed in \cite{CarmonaandDurrleman2003a}, we have
\begin{equation}\label{parity}
\Pi_T(F_1, \sigma_1, F_2, \sigma_2, K, \rho)=\Pi_T(F_2, \sigma_2, F_1, \sigma_1, -K, \rho)+e^{-rT}(F_1-F_2-K).
\end{equation}
This shows that a closed-form formula for (\ref{spreadprice}) when $K>0$ can be used to obtain a closed-form formula for (\ref{spreadprice}) when $K<0$ by using (\ref{parity}). Therefore in our discussions below we assume $K>0$. We also drop the measure $Q$ when we take expectations. All the expectations below are understood under the risk-neutral measure $Q$. 

 Denote by $(Y, X)$ a two dimensional Gaussian random vector with the same distribution as $\frac{1}{\sqrt{T}}(W_1(T), W_2(T))$. Then both $X$ and $Y$ are standard normal and $Cov(X, Y)=\rho$. With these new notations, we can express the spread option price as
\begin{equation}\label{6}
\Pi_T=e^{-rT}E[(F_1e^{-\frac{1}{2}\sigma^2_1T+\sigma_1\sqrt{T}Y}-F_2e^{-\frac{1}{2}\sigma^2_2T+\sigma_2\sqrt{T}X} -K)^+].    \end{equation}
It can be easily seen from (\ref{6}) above that when $F_2=0$ the spread price $\Pi_T$ is given by Black-Scholes formula \citep{black1976pricing}. When $K=0$, the spread option becomes an exchange option and in this case we have  closed form formula, see \cite{margrabe1978value} for this. For general model parameters, there is no exact closed-form solution exits
for $\Pi_T$. As mentioned in the introduction, a number of papers have presented closed form approximation formulas for the spread prices $\Pi_T$ in the past.

Our goal in this paper is to obtain an alternative approximate closed-form formula for the above equation (\ref{6}). To achieve this goal, we first express the spread price by the probabilities of three events and then approximate these three probabilities separately to obtain an approximate closed form formula for $\Pi_T$.

Throughout the paper we use the following notations. For any $x=(x_1, x_2, x_3, x_4, x_5, \rho)$ we write
\begin{equation}\label{666}
\Pi_T(x)=:e^{-rT}E[(x_1e^{-\frac{1}{2}x^2_2T+x_2\sqrt{T}Y}-x_3e^{-\frac{1}{2}x^2_4T+x_4\sqrt{T}X} -x_5)^+].    
\end{equation}
We denote 
\begin{equation}\label{theta}
\theta=:(F_1, \sigma_1, F_2, \sigma_2, K, \rho), \; \;    \bar{\theta}=:(F_2, \sigma_2, F_1, \sigma_1, -K, \rho). 
\end{equation}
With these notations the put-call parity relation (\ref{parity}) is written as
\begin{equation}
\Pi_T(\theta)=\Pi_T(\bar{\theta})+e^{-rT}(F_1-F_2-K).    
\end{equation}
We use $\varphi$ to denote the probability density function of the standard normal random variable and by $\Phi$ we denote the cumulative distribution function of the standard normal random variable.
\section{Carmona-Durrleman and Bjerksund-Stensland formulas}

The main purpose of this section is to give a review for the approaches in the two noticeable papers \cite{CarmonaandDurrleman2003a}
and \cite{bjerksund2014closed} in pricing spread options under log-normal models. In our Proposition \ref{41} below, we explain that the Carmona-Durrleman formula always gives higher spread option prices than the Bjerksund-Stensland formula. This, both of them being lower bounds to the true spread prices, shows that the Carmona-Durrleman formula is more accurate than the Bjerksund-Stensland formula.

First recall that we are interested in computing $EG^+$, where 
\begin{equation}\label{G}
G= F_1e^{-\frac{1}{2}\sigma^2_1T+\sigma_1\sqrt{T}Y}-F_2e^{-\frac{1}{2}\sigma^2_2T+\sigma_2\sqrt{T}X} -K, \end{equation}
and $Y, X$ are standard normal random variables with $Cov(Y, X)=\rho$, see (\ref{6}) and the arguments preceding to it for the details of this. The paper  \cite{CarmonaandDurrleman2003a} gives lower bounds for $EG^+$ by using the arguments in their Proposition 2.  In their paper they write down $Y$ , adapting to our own notation,  as $Y=Z\sin \phi+X\cos \phi$, where $Z$ is a standard normal random variable that is independent from $X$ and $\phi$ is such that $\cos\phi=\rho$. They introduce a family of random variables of the form $Y_{\theta}=Z\sin \theta+X\cos \theta$ and calculate
\begin{equation}\label{RC0}
\bar{\Pi}_T^{CD}(\theta, d)=: E[(S_1(T)-S_2(T)-K)I(Y_{\theta}\le d)].   
\end{equation}
Then the spread option price $\Pi_T$ in (\ref{6}) is approximated by $e^{-rT}\hat{\Pi}_T^{CD}$ with
\begin{equation}\label{RC1}
\hat{\Pi}_T^{CD}=:\sup_{\theta \in R }\sup_{d\in R}\bar{\Pi}_T^{CD}(\theta, d).
\end{equation}

A simple application of Girsanov theorem (see the proof in Appendix B of \cite{bjerksund2014closed}) gives
\begin{equation}\label{RC2}
\bar{\Pi}_T^{CD}(\theta, d)
=F_1\Phi(d+\sigma_1\sqrt{T} \cos(\theta+\phi))-F_2\Phi(d+\sigma_2\sqrt{T} \cos \theta)-K\Phi(d),\\   
\end{equation}
where $\Phi(\cdot)$ denotes the cumulative distribution function of a standard normal random variable. So the problem of finding (\ref{RC1}) reduces to maximizing the function on the right hand side of (\ref{RC2}) for $\theta$ and $d$. The first order conditions for maximization gives the following two equations
\begin{equation}
 \begin{split}
 &F_1\varphi(d+\sigma_1 \sqrt{T} \cos(\theta+\phi))-F_2\varphi(d+\sigma_2\sqrt{T} \cos \theta)-K\varphi(d)=0,\\
 &F_1\sigma_1\sqrt{T}\sin(\theta+\phi)\varphi(d+\sigma_1 \sqrt{T}\cos(\theta+\phi))-F_2\sigma_2 \sqrt{T}\sin \theta \varphi(d+\sigma_2 \sqrt{T}\cos \theta)+K\varphi(d)=0,\\
  \end{split}   
\end{equation}
where $\varphi$ denotes the probability density function of a standard normal random variable. These two equations can be solved for $\varphi(d+\sigma_1\sqrt{T} \cos(\theta+\phi))$ 
and $\varphi(d+\sigma_2\sqrt{T} \cos \theta)$ as follows
\begin{equation}\label{max}
\begin{split}
 \varphi(d+\sigma_1\sqrt{T} \cos(\theta+\phi))&=\frac{\sigma_2\sqrt{T}\sin\theta+1}{\sigma_2\sqrt{T}\sin\theta -\sigma_1\sqrt{T}\sin(\theta+\phi)}\frac{K\varphi(d)}{F_1},\\
 \varphi(d+\sigma_2\sqrt{T} \cos \theta)&=\frac{1+\sigma_1
 \sqrt{T}\sin(\theta+\phi)}{\sigma_2\sqrt{T}\sin\theta-\sigma_1\sqrt{T}\sin(\theta+\phi)}\frac{K\varphi(d)}{F_2}.
 \end{split}
\end{equation}

Denoting the solution of (\ref{max}) by $\theta^{\star}$ and $d^{\star}$, the spread option price presented in \cite{CarmonaandDurrleman2003a} equals to
\begin{equation}\label{CD-1}
 \Pi_T^{CD}=e^{-rT} F_1\Phi(d^{\star}+\sigma_1 \sqrt{T} \cos(\theta^{\star}+\phi))-e^{-rT}F_2\Phi(d^{\star}+
 \sigma_2 \sqrt{T}\cos \theta^{\star})-e^{-rT}K\Phi(d^{\star}).  
\end{equation}
Note here that the equation (\ref{max}) above is equivalent to the equation (12) in the paper \cite{CarmonaandDurrleman2003a}. 

The above approach provides a method in pricing spread options by constructing proper lower bounds to it and then optimizing these lower bounds. The Carmona-Durrleman formula gives very precise spread price. However, to obtain spread price one needs to solve the equations (\ref{max}) for $d^{\star}$ and $\theta^{\star}$ numerically. The paper \cite{CarmonaandDurrleman2003a} in fact reduced this problem further and one only needs to solve $\theta^{\star}$ numerically from (13) in their paper and use the relation for $d^{\star}$ in Proposition 6 in their paper to obtain $d^{\star}$. While this method reduces the problem a bit, one still needs numerical procedure to obtain the prices of spread options.

Another interesting procedure in pricing spread options were proposed in the recent paper \cite{bjerksund2014closed}. The paper also introduced  lower bounds for the spread call option prices as follows 
\begin{equation}\label{BS1}
\Pi_T^{BS}(a, b)=e^{-rT}E\big [(S_1(T)-S_2(T)-K)I\big ( S_1(T)\geq \frac{a(S_2(T))^b}{E[(S_2(T))^{b}]}\big )\big ],
\end{equation}
where $a$ and $b$ are two parameters that needs to be determined by optimizing the lower bounds  $\Pi_T^{BS}(a, b)$. Their proposed approximate spread option price is given by
\begin{equation}
\Pi_T^{BS}=\sup_{a, b}  \Pi_T^{BS}(a, b).  
\end{equation}
In their paper the authors set
\begin{equation}\label{ab}
a=F_2+K, \;\;\; b=\frac{F_2}{F_2+K}.    
\end{equation}

By evaluating the equation (\ref{BS1}), they obtained the following formula for the spread call option price
\begin{equation}\label{BSS1}
\Pi_T^{BS}=e^{-rT}[F_1\Phi(\bar{d}_1)-F_2\Phi(\bar{d}_2)-K\Phi(\bar{d}_3)],    
\end{equation}
where 
\begin{equation}\label{BSS2}
\begin{split}
\bar{d}_1&=\frac{\ln\left( \frac{F_1}{a}\right)+(\frac{1}{2}\sigma_1^2+\frac{1}{2}b^2\sigma_2^2-b\rho \sigma_1\sigma_2 )T} {\sigma\sqrt{T}},\\
\bar{d}_2&=\frac{\ln\left( \frac{F_1}{a}\right)+(-\frac{1}{2}\sigma_1^2-b\sigma_2^2+\frac{1}{2}b^2\sigma_2^2+\rho \sigma_1\sigma_2)T} {\sigma\sqrt{T}},\\
\bar{d}_3&=\frac{\ln\left( \frac{F_1}{a}\right)+(-\frac{1}{2}\sigma_1^2+\frac{1}{2}b^2\sigma_2^2)T} {\sigma\sqrt{T}},\\
\end{split}    
\end{equation}
and
\begin{equation}\label{BSS3}
 \sigma= \sqrt{\sigma_1^2+b^2\sigma_2^2-2\rho b\sigma_1\sigma_2}.  
\end{equation}
 
 Clearly, the Bjerksund-Stensland formula is simpler than the Carmona-Durrleman formula as one does not have to apply numerical procedure to obtain spread option price in the Bjerksund-Stensland formula. However, as mentioned earlier, the Carmona-Durrleman formula gives more precise spread option prices than the Bjerksund-Stensland formula. 
 
 To see this, we write the relation $S_1(T)\geq \frac{a(S_2(T))^b}{E[(S_2(T))^{b}]}$ in (\ref{BS1}) by using $Y_{\theta}$ defined as in the paper \cite{CarmonaandDurrleman2003a}. We  obtain that in fact the lower bounds $\Pi_T^{BS}(a, b)$ are a subset of the lower bounds $\Pi_T^{CD}(\theta, d)$ proposed by \cite{CarmonaandDurrleman2003a}. We state this fact as a proposition and write down the proof in the appendix.
\begin{proposition}\label{41} We have
\begin{equation}
\Pi_T^{BS}(a, b)=e^{-rT}\bar{\Pi}_T^{CD}(\theta_0, d_0),    \end{equation}
where the parameters $\theta_0$ and $d_0$ are given by the following relations
\begin{equation}
\sin\theta_0=\frac{-\sigma_1 \sin\phi}{\sqrt{\sigma_1^2+\sigma_2^2b^2-2\sigma_1\sigma_2b\cos \phi}},\;\; \cos \theta_0= \frac{-\sigma_2b+\sigma_1 \cos\phi}{\sqrt{\sigma_1^2+\sigma_2^2b^2-2\sigma_1\sigma_2b\cos \phi}},  
\end{equation}
and
\begin{equation}
 d_0=\frac{\ln \frac{F_1}{a}-\frac{1}{2}\sigma_1^2T+\frac{1}{2}b^2\sigma_2^2T}{\sqrt{\sigma_1^2+\sigma_2^2b^2-2\sigma_1\sigma_2b\cos \phi}\sqrt{T}}.  
\end{equation}
This shows the following relation
\begin{equation}\label{ineq}
\Pi_T^{BS}\le \Pi_T^{CD}.    
\end{equation}
\end{proposition}

\begin{remark} We remark that the relation (\ref{ineq}) shows that the Carmona-Durrleman spread option price is more accurate than the Bjerksund-Stensland spread option price. The advantage of the Bjerksund-Stensland formula over the Carmona-Durrleman formula however is that it is in closed-form and one does not need numerical procedure to obtain the spread option price. \end{remark}
\begin{remark} We remark here that it was brought to our attention by the referees that
the result of our Proposition \ref{41} above was also discussed in Appendix F of \cite{bjerksund2014closed}. Our current proposition expresses this relation (\ref{ineq}) more explicitly only. 
\end{remark}

\section{A numerical procedure}
In this section, we first express the spread price (\ref{spreadprice}) as a linear combination of the probabilities of three events, see Proposition \ref{prop1} below. Such representation will help us to derive a closed form spread option valuation formula in the next section, see Proposition \ref{mainmain}. We also  write down a numerical approximation formula (see Proposition \ref{prop2} below) of the spread price by using  Simpson's rule for Riemann integrals. At the end of this section, we compare our formula (\ref{C})  with a related formula discussed in \cite{Ravindran}. 

The following proposition  expresses the price of the spread option by probabilities of three events under the  risk-neutral  measure. The proof of this result uses the Girsanov's change of measure technique for log-normal models, see the proof in Appendix C. 
\begin{proposition}\label{prop1} The price of the spread option is given by
\begin{equation}\label{C} 
\Pi_T=e^{-rT}F_{1}C_D^1-e^{-rT}F_{2}C_D^2-e^{-rT}KC_D^3, 
\end{equation}
where
\begin{equation*}
\begin{split}
C_D^1&=Q\left(g_1\bar{F}_{1}e^{\sigma_{1}\sqrt{T}Y}-\alpha \bar{F}_{2}e^{\sigma_{2}\sqrt{T}X}-K\geq 0 \right),\\
C_D^2&=Q\left(\alpha \bar{F}_{1}e^{\sigma_{1}\sqrt{T}Y}-g_2\bar{F}_{2}e^{\sigma_{2}\sqrt{T}X}-K\geq0\right),\\
C_D^3&=Q\left(\bar{F}_{1}e^{\sigma_{1}\sqrt{T}Y}-\bar{F}_{2}e^{\sigma_{2}\sqrt{T}X}-K\geq 0\right),\\
\end{split}
\end{equation*}
and  $\alpha=e^{\rho \sigma_1\sigma_2 T}$, $g_1=e^{\sigma_{1}^{2}T}$, $g_2=e^{\sigma_{2}^{2}T}$, $\bar{F}_1=F_1e^{-\frac{1}{2}\sigma_1^2T}$, $\bar{F}_2=F_2e^{-\frac{1}{2}\sigma_2^2T}$. 
\end{proposition}

We remark that $C_D^1$ above represents the probability (under the risk neutral measure $Q$) that a call option with strike price $K$ and maturity $T$ on the spread between $g_1$ stocks with price dynamics $S_1(t)$  and $\alpha$ stocks with price dynamics $S_2(t)$ is exercised (ends up in the money). $C_D^2$ can be interpreted as the exercise probability of a call on the spread between $\alpha$ stocks with price dynamics $S_1(t)$ and $g_2$ stocks with price dynamics 
$S_2(t)$. $C_D^3$ can be interpreted as the exercise probability of the call option with strike price $K$ of the spread $S_1(T)-S_2(T)$.

\let\thefootnote\relax\footnote{The formula (\ref{C}) has a simple form when $K=0$. In fact, when $K=0$, the formula (\ref{C}) reduces to Margrabe's formula \cite{margrabe1978value}: 
\begin{equation*}
\Pi_T^M=e^{-rT}F_1\Phi \left (\frac{\ln\left( \frac{g_1\bar{F}_1}{\alpha \bar{F}_2}\right)}{\sqrt{T}\sqrt{\sigma_1^2+\sigma_2^2-2\sigma_2\sigma_2\rho}}\right )-e^{-rT}F_2\Phi \left ( \frac{\ln\left( \frac{\alpha \bar{F}_1}{g_2 \bar{F}_2}\right)}{\sqrt{T}\sqrt{\sigma_1^2+\sigma_2^2-2\sigma_2\sigma_2\rho}}   \right )   \end{equation*}
}
Next, in Proposition \ref{prop2} below, we write down a numerical procedure  that approximates the spread price. Before we state our next result, we first fix some notations. Being a jointly Gaussian random variable with $Cov(Y, X)=\rho$, $(Y, X)$ has the property that $X\sim N(0, 1)$ and $Y|X=a \sim N(a\rho, 1-\rho^2)$. We use this fact to write down an approximate formula for the spread call option price below. Let $b>0$ be a number that the probability of the event $X\in [-b, b]^c$ is very small. Here the notation $c$ represents the complement of an event. We divide the interval $[-b, b]$ into small equally spaced $N$ intervals for a large integer number $N>0$. We denote $a_i=b(\frac{2i}{N}-1)$ for $i=1, 2, \cdots, N$. With these notations fixed, we state our next result

\begin{proposition}\label{prop2} For $K>0$,  let
\begin{equation}\label{Ga}
G(a)=:F_1\Phi(d^1(a))-F_2\Phi(d^2(a))-K\Phi(d^3(a)),
\end{equation}
where
\begin{equation}
\begin{split}
d^1(a)&=\frac{-\ln\left(\frac{\alpha \bar{F}_2}{g_1\bar{F}_1}e^{\sigma_2 \sqrt{T}a}+\frac{K}{g_1\bar{F}_1}\right)
+\sigma_1 \rho \sqrt{T} a}{\sigma_1 \sqrt{T}\sqrt{1-\rho^{2}}},\\
d^2(a)&=\frac{-\ln\left(\frac{g_2\bar{F}_2}{\alpha \bar{F}_1}e^{\sigma_2 \sqrt{T}a}+\frac{K}{\alpha \bar{F}_1}\right)+
\sigma_1 \rho \sqrt{T} a}{\sigma_1 \sqrt{T}\sqrt{1-\rho^{2}}},\\
d^3(a)&=\frac{-\ln\left(\frac{\bar{F}_2}{\bar{F}_1}e^{\sigma_2 \sqrt{T}a}+\frac{K}{\bar{F}_1}\right)
+\sigma_1 \rho \sqrt{T} a}{\sigma_1 \sqrt{T}\sqrt{1-\rho^{2}}},\\
\end{split}
\end{equation}
with  $\bar{F}_1$, $\bar{F}_2$, $\alpha, g_1, g_2,$ given by Proposition \ref{prop1} above.
Then we have $\Pi_T=e^{-rT}\bar{\Pi}_T$ with 
\begin{equation}\label{new42}
\bar{\Pi}_T=\int_{-\infty}^{+\infty}G(a)d\Phi(a)=\int_{-\infty}^{+\infty}G(a)\varphi(a)da.    
\end{equation}
Denoting $f(a)=:G(a)\varphi(a)$,  the $\bar{\Pi}_T$ in (\ref{new42}) can be approximated by $\bar{\Pi}_T^d(N, b)$ given by
\begin{equation}\label{266S}
 \bar{\Pi}_T^d(N, b)=:\frac{2b}{3N}[f(b)+f(-b)]+\frac{8b}{3N}
\sum_{i=1}^{N-1}\frac{1+(-1)^i}{2}f(a_i)+\frac{4b}{3N}
\sum_{i=1}^{N-1}\frac{1+(-1)^{i+1}}{2}f(a_i),
\end{equation}
for properly chosen positive integer $N$ and positive number $b$, where for each $i=0, 1, \cdots, N$,  $a_{i}=b(\frac{2i}{N}-1)$.
\end{proposition}

\begin{remark} \label{Simp} In the above Proposition  \ref{prop2} we  assume $K>0$. If $K<0$, we use (\ref{parity}). When $K=0$ the spread price $\Pi_T$ is given by Margrabe's formula as discussed earlier. 
\end{remark}
\begin{remark}
The approximation (\ref{266S}) is obtained by applying Simpson's rule for the Riemann integral $\int_{-b}^bG(a)\varphi(a)da$. By using the well known error bound for Simpson's rule, we obtain
\begin{equation}\label{SM}
|\bar{\Pi}_T-\bar{\Pi}_T^d(N, b)|\le  \frac{M(2b)^5}{180N^4},   
\end{equation}
where $M$ is the maximum value of $f^{(4)}(a)$ (the fourth order derivative of $f(a)=G(a)\varphi(a)$) on $[-b, b]$. Clearly, the speed and precision of the approximation (\ref{266S}) depends on the model parameters. The $M$ in (\ref{SM}) depends on the model parameters also
and it controls the error bound. Obtaining an explicit expression for 
 $M$ is not trivial. However our extensive numerical tests show that $e^{-rT}\bar{\Pi}_T^d(500, 5)$ gives pretty good approximation for spread option prices for most of the  model parameters.
\end{remark}

\begin{remark} One can also approximate the Riemann-Stieltjes integral in (\ref{new42}) as
\begin{equation}\label{266}
 \bar{\Pi}_T^{RS}(N, b)=:\sum_{i=0}^{N-1}[\frac{G(a_{i+1})+G(a_i)}{2}][\Phi(a_{i+1})-\Phi(a_{i})],
\end{equation}
for properly chosen positive integer $N$ and positive number $b$, where for each $i=0, 1, \cdots, N$,  $a_{i}=b(\frac{2i}{N}-1)$, and use $\bar{\Pi}_T^{RS}(N, b)$ as an approximation formula for the spread price. This approximation (\ref{266}) is trapezoidal quadrature rule for Riemann-Stieltjes integral, see \citep{Dragomir} for example. By Theorem 8 of \cite{Dragomir}, we have the following error bound
\[
|\bar{\Pi}_T-\bar{\Pi}_T^{RS}(N, b)|\le \frac{Hb}{N}[\Phi(b)-\Phi(-b)],
\]
where $H=max_{a\in [-b, b]}|G'(a)|$. It can be easily checked that when $K>0$ we have
\begin{equation}\label{GKU}
max_{x\in R}|G'(x)|\le \frac{1}{\sqrt{2\pi (1-\rho^2)}\sigma_1}(F_1+F_2+K)(\sigma_2+\sigma_1\rho),
\end{equation}
and when $K<0$, by put-call parity, we have
\begin{equation}\label{GKD}
max_{x\in R}|G'(x)|\le \frac{1}{\sqrt{2\pi (1-\rho^2)}\sigma_2}(F_1+F_2-K)(\sigma_1+\sigma_2\rho).
\end{equation}
Our numerical tests show, however, that (\ref{266}) does not perform as good as the approximation (\ref{266S}) obtained by the Simpson's rule.
\end{remark}

Another relevant approach to our Proposition \ref{prop2} above is given in the paper \cite{Ravindran}. Here we review this approach. Denoting by $E_Z$ the expectation with respect to a random variable $Z$, the paper \cite{Ravindran} writes the spread price $\Pi_T$ in (\ref{6}) as 
\begin{equation}\label{277}
\Pi_T^R=e^{-rT}E_X\Big [E_Y\big [(F_1e^{-\frac{1}{2}\sigma^2_1T+\sigma_1\sqrt{T}Y}-F_2e^{-\frac{1}{2}\sigma^2_2T+\sigma_2\sqrt{T}X} -K)^+/X\big ]\Big ].
\end{equation}
The inner expectation  $E_Y[\big (F_1e^{-\frac{1}{2}\sigma^2_1T+\sigma_1\sqrt{T}Y}-F_2e^{-\frac{1}{2}\sigma^2_2T+\sigma_2\sqrt{T}X} -K)^+/X\big ]$ is evaluated first and then the outer expectation $E_X$ is found. This approach leads to an exact formula for the spread price $\Pi_T$ as in  \cite{Ravindran}. The expression (\ref{277}) can be simplified to (\ref{RRR}) (adapted to our notation) in the footnote below. The derivation will be given in Appendix C.

\let\thefootnote\relax\footnote{The expressions in (\ref{277}) can be simplified  to: \begin{equation}\label{RRR} 
\Pi^R=e^{-rT}E[\tilde{F}(X)\Phi(\tilde{d}_1(X))]-e^{-rT}E[\tilde{K}(X)\Phi(\tilde{d}_2(X))],    
\end{equation}
where 
\begin{equation*}
\tilde{d}_{1}(x)=1/\tilde{\sigma}\sqrt{T}\left(\ln\left(\tilde{F}(x)/\tilde{K}(x)\right)+\tilde{\sigma}^{2}T/2\right),\;
 \tilde{d}_{2}(x)=1/\tilde{\sigma}\sqrt{T}\left(\ln\left(\tilde{F}(x)/\tilde{K}(x)\right)-\tilde{\sigma}^{2}T/2\right),
\end{equation*}
and 
\[
\tilde{F}(x)=F_{1}e^{-\frac{1}{2}\sigma_{1}^{2}\rho^{2}T+\sigma_{1}\sqrt{T}\rho x},\; \tilde{K}(x)=F_{2}e^{-\frac{1}{2}\sigma_{2}^{2}T+\sigma_{2}\sqrt{T}x}+K,\; \tilde{\sigma}=\sigma_{1}\sqrt{1-\rho^{2}}.
\]
}
The formula (\ref{RRR}) can be evaluated by using the above mentioned trapezoidal rule for Riemann-Stieltjes integrals or by using other Gaussian quadrature methods. 

\begin{remark}\label{rem4.44} We remark that the relation (\ref{C}) can also be derived by using (\ref{RRR}). The proof of this is given in the Appendix C. This shows, in particular, that (\ref{C}) can be obtained by conditioning as it was done in \cite{Ravindran} and without relying on Girsanov's change of measure theorem. 
\end{remark}

\section{Generalization of the Bjerksund-Stensland formula}
As mentioned earlier, the \cite{bjerksund2014closed} formula presented in Section 3 above performs highly accurately. In this section, we study the Bjerksund and Stensland formula further and give an alternative derivation for it. Our new approach in deriving the Bjerksund and Stensland formula in this section leads us to a more general formula than the Bjerksund and Stensland formula as our Corollary \ref{cor52} below shows.

As pointed out in Section 4 above, the \cite{CarmonaandDurrleman2003a} and \cite{bjerksund2014closed} spread option prices were obtained by maximizing lower bounds for the spread prices. In this section, instead of constructing lower bounds for spread call option prices, we directly approximate the probabilities $C_D^1, C_D^2, C_D^3$ in the Proposition \ref{prop1} above. 

Before presenting our main result of this paper, we first define the following three curves in the $x$-$y$ coordinate system:
\begin{equation}\label{curves}
\begin{split}
\mathcal{C}_1:&\; \;  g_1\bar{F}_{1}e^{\sigma_{1}\sqrt{T}y}-\alpha \bar{F}_{2}e^{\sigma_{2}\sqrt{T}x}-K=0, \\
\mathcal{C}_2:&\; \;  \alpha \bar{F}_{1}e^{\sigma_{1}\sqrt{T}y}-g_2 \bar{F}_{2}e^{\sigma_{2}\sqrt{T}x}-K=0,\\
 \mathcal{C}_3:&\;\; \bar{F}_{1}e^{\sigma_{1}\sqrt{T}y}- \bar{F}_{2}e^{\sigma_{2}\sqrt{T}x}-K=0.\\  
\end{split}
\end{equation}\label{line}
We denote by $\mathcal{C}_1^{+}, \mathcal{C}_2^{+}, \mathcal{C}_3^{+}$ the domains that lie above the three curves $\mathcal{C}_1,\mathcal{C}_2,\mathcal{C}_3$ in the $x$-$y$ coordinate system respectively. Namely, the domain $\mathcal{C}_1^+$, for example, is given by all  $(x, y)$ such that
\begin{equation}
 g_1\bar{F}_{1}e^{\sigma_{1}\sqrt{T}y}-\alpha \bar{F}_{2}e^{\sigma_{2}\sqrt{T}x}-K\geq 0,  
\end{equation}
and the other two domains $\mathcal{C}_2^+$ and $\mathcal{C}_3^+$ are defined similarly.

Our approach in this section is to approximate the following probabilities 
\begin{equation}\label{prob}
  Q((X, Y)\in \mathcal{C}_i^+)  
\end{equation}
in closed form for each $i=1, 2, 3$. 

The difficulty in obtaining closed form approximations for (\ref{prob}) above  is  that the boundaries $\mathcal{C}_1, \mathcal{C}_2, \mathcal{C}_3$ of the domains $\mathcal{C}_1^+, \mathcal{C}_2^+, \mathcal{C}_3^+ $ are not linear functions. Therefore, we wish to find  linear functions $y=\kappa_ix+\delta_i$
with appropriate slope $\kappa_i$ and intersection $\delta_i$ for each $i=1,2,3,$ so that the following approximations hold
\begin{equation}\label{approx1}
  Q((x, y)\in \mathcal{C}_i^+)\approx Q(Y\geq \kappa_i X+\delta_i), i=1,2, 3.   
\end{equation}

If we can achieve (\ref{approx1}), then the latter probabilities in (\ref{approx1}) can be written in closed-form by using the following lemma
\begin{lemma} \label{lem1} For any two standard normal random variables $Y$ and $X$ with $Cov(Y, X)=\rho$, and any three real numbers $m, n, \ell,$ we have
\begin{equation}
Q(mY-nX\geq \ell)=\Phi(\frac{-\ell}{\sqrt{m^2+n^2-2\rho mn}} )    
\end{equation}
\end{lemma}

The application of Lemma \ref{lem1}  above leads us to
\begin{equation}\label{approx2}
  Q((x, y)\in \mathcal{C}_i^+)\approx Q(Y\geq \kappa_i X+\delta_i)=\Phi\big (\frac{-\delta_i}{\sqrt{1+\kappa_i^2-2\rho \kappa_i}}\big), i=1,2,3,
 \end{equation}
and this will give us to a closed-form formula for the spread option price due to (\ref{C}).

To construct lines $y=\kappa_i x+\delta_i$ that makes the above approximation as precise as possible, we  fix some point $(x_0^i, y_0^i)$ on the curve $\mathcal{C}_i$ and some slope $\kappa_i$ for each $i=1,2,3,$ and use the following lines
\begin{equation}\label{approx33}
 y=\kappa_i(x_0)x+\delta_i(x_0),\;\;  \delta_i(x_0)=y_0-\kappa_i(x_0)x_0,\; i=1, 2, 3.
\end{equation}

In the next lemma, we state some properties of the curves $\mathcal{C}_1, \mathcal{C}_2, \mathcal{C}_3$ above. The property (a) of this lemma is useful for our discussions in the remaining of this paper.
\begin{lemma}\label{lem2} The three curves $\mathcal{C}_1, \mathcal{C}_2, \mathcal{C}_3$ have the following properties
\begin{enumerate}
\item[(a)] For all the three curves $\mathcal{C}_1, \mathcal{C}_2, \mathcal{C}_3$ we have
\begin{equation*}
\lim_{x\rightarrow +\infty}\frac{y}{x}=\frac{\sigma_2}{\sigma_1}.
\end{equation*}
\item[(b)] When $x\rightarrow -\infty$, the limit $\lim_{x\rightarrow -\infty} y$ exists and  
\begin{equation*}
\lim_{x\rightarrow -\infty}y=\left \{
\begin{array}{ll}
\frac{1}{\sigma_1\sqrt{T}}\ln\left(\frac{K}{\bar{F}_1}\right)-\sigma_1 \sqrt{T}&\mbox{for $\mathcal{C}_1$},\\
\frac{1}{\sigma_1\sqrt{T}}\ln \left(\frac{K}{\bar{F}_1}\right)-\rho \sigma_2 \sqrt{T}&\mbox{for $\mathcal{C}_2$},\\
\frac{1}{\sigma_1\sqrt{T}}\ln \left(\frac{K}{\bar{F}_1}\right)&\mbox{for $\mathcal{C}_3$}.\\
\end{array}
\right.
\end{equation*}
\end{enumerate}
\end{lemma}
Part (a) of the above Lemma \ref{lem2} shows that the slopes of the asymptotic lines of all the three curves $\mathcal{C}_1, \mathcal{C}_2, \mathcal{C}_3$ when $x\rightarrow +\infty$ is the same and is given by $\frac{\sigma_2}{\sigma_1}$. Part (b) gives lower bounds for each curves when $x\rightarrow -\infty$.

\subsection{Alternative derivation of the Bjerksund-Stensland formula}
The purpose of this subsection is to derive the Bjerksund and Stensland spread option formula presented in \cite{bjerksund2014closed}  by using the arguments discussed in  (\ref{approx1}), (\ref{approx2}), (\ref{approx33}) above. In the next subsection, we apply a similar approach and obtain a generalization of the \cite{bjerksund2014closed} spread option formula.

We consider two points $(x_0^i, y_0^i)$ and $(x_{\ell}^i, y_{\ell}^i)$ on the curve $\mathcal{C}^i$ for each $i=1, 2, 3$. The equation of the line that passes through these two points are given by 
\begin{equation*}
y^i=\frac{y_{\ell}^i-y_0^i}{x_{\ell}^i-x_0^i}x^i+y_0^i-\frac{y_{\ell}^i-y_0^i}{x_{\ell}^i-x_0^i}x_0^i, i=1,2,3.
\end{equation*}
From part (a) of Lemma \ref{lem2}, we see that when $x_{\ell}^i$ goes to $+\infty$,  $\frac{y_{\ell}^i-y_0^i}{x_{\ell}^i-x_0^i}$ monotonically increases to $\frac{\sigma_2}{\sigma_1}$. So we  introduce a parameter $0<b\le 1$ and define the following lines
\begin{equation}\label{1s}
 \ell^i: y^i=b\frac{\sigma_2}{\sigma_1}x^i+y_0^i-b\frac{\sigma_2}{\sigma_1}x_0^i, i=1,2,3.   
\end{equation}
Our goal is to use these lines $\ell_i$ in (\ref{1s}) to obtain a closed-form formula for the spread option. For this we need to discuss how to choose appropriate points $(x_0^i, y_0^i)$. Since $(x_0^i, y_0^i)$ lies on the curve $\mathcal{C}_i$ for each $i=1,2,3$, we plug in these points to the equations of the three curves and obtain the expressions for $y_0^i, i=1,2,3,$ 
\begin{equation}\label{ys}
\begin{split}
y_0^1&=\frac{1}{\sigma_1 \sqrt{T}}\ln \left(\frac{F_2e^{\rho \sigma_1 \sigma_2 T+ (\sigma_2\sqrt{T}x_0^1-\frac{1}{2}\sigma_2^2T)}+K}{F_1}\right)-\frac{\sigma_1\sqrt{T}}{2},\\ 
y_0^2&=\frac{1}{\sigma_1 \sqrt{T}}\ln\left( \frac{F_2e^{\sigma_2^2T+(\sigma_2\sqrt{T}x_0^2-\frac{1}{2}\sigma_2^2T)} +K}{F_1}\right)+(\frac{1}{2}\sigma_1-\rho \sigma_2)\sqrt{T},\\  
y_0^3&=\frac{1}{\sigma_1 \sqrt{T}}\ln\left( \frac{F_2e^{\sigma_2\sqrt{T}x_0^3-\frac{1}{2}\sigma_2^2T}+K}{F_1}\right)+\frac{\sigma_1\sqrt{T}}{2}.\\  
\end{split}    
\end{equation}

To simplify the expressions for $y_0^i, i=1, 2, 3,$ above we apply the following approximations
\begin{equation}\label{yapprox}
\begin{split}
 F_2e^{\rho \sigma_1 \sigma_2 T+ (\sigma_2\sqrt{T}x_0^1-\frac{1}{2}\sigma_2^2T)}+K&\approx a_1 e^{\rho \sigma_1q_1T+q_1\sqrt{T}x_0^1  -\frac{1}{2}q^2_1T},\\   
 F_2e^{\sigma_2^2T+(\sigma_2\sqrt{T}x_0^2-\frac{1}{2}\sigma_2^2T)} +K&\approx a_2e^{ q_2\sigma_2T+q_2\sqrt{T}x_0^2-\frac{1}{2}q^2_2T},\\
 F_2e^{\sigma_2\sqrt{T}x_0^3-\frac{1}{2}\sigma_2^2T}+K&\approx a_3e^{q_3\sqrt{T}x_0^3-\frac{1}{2}q^2_3T}, \\
\end{split}    
\end{equation}
for some appropriate $a_1, a_2, a_3$ and $q_1, q_2, q_3$. We remark that the approximations (\ref{yapprox}) are similar to Kirk approximation that was discussed in Appendix A. For example, one can let 
\begin{equation}\label{aq}
a_1=a_2=a_3=F_2+K, q_1=q_2=q_3=\frac{F_2}{F_2+K}\cdot\sigma_2
\end{equation}
as in Kirk approximation. 

We would like to implement the approximations (\ref{yapprox}) to  (\ref{ys}) and obtain approximated values for these points $y_0^1, y_0^2, y_0^3$. To this end, we define the following functions
\begin{equation}\label{barys}
\begin{split}
z_0^1(q, a, h)&=\frac{1}{\sigma_1 \sqrt{T}}\ln \left(\frac{ae^{\rho \sigma_1qT+q\sqrt{T}h  -\frac{1}{2}q^2T}}{F_1}\right)-\frac{\sigma_1\sqrt{T}}{2},\\  
z_0^2(q, a, h)&=\frac{1}{\sigma_1 \sqrt{T}}\ln\left( \frac{ae^{ q\sigma_2T+q\sqrt{T}h-\frac{1}{2}q^2T} }{F_1}\right)+(\frac{1}{2}\sigma_1-\rho \sigma_2)\sqrt{T},\\ 
z_0^3(q, a, h)&=\frac{1}{\sigma_1 \sqrt{T}}\ln\left( \frac{ae^{q\sqrt{T}h-\frac{1}{2}q^2T}}{F_1}\right)+\frac{\sigma_1\sqrt{T}}{2}.\\  
\end{split}    
\end{equation}
Then with (\ref{yapprox}) and (\ref{aq}), we have
\begin{equation}
\begin{split}
y_0^1&\approx z_0^1(\frac{F_2}{F_2+K}\cdot\sigma_2, F_2+K, x_0^1), \;\;\;  y_0^2\approx z_0^2(\frac{F_2}{F_2+K}\cdot\sigma_2, F_2+K, x_0^2),\\
y_0^3&\approx z_0^3(\frac{F_2}{F_2+K}\cdot\sigma_2, F_2+K, x_0^3). 
\end{split}
\end{equation}

Now, we are ready to state our next result. The following proposition shows that the spread option price $\Pi^{BS}$ presented in \cite{bjerksund2014closed} can also be obtained by using the ideas in relations (\ref{approx1}) and (\ref{approx2}) above.
\begin{proposition} \label{BS-formula}  We have
\begin{equation}\label{BS11}
\begin{split}
 \Pi^{BS}&=e^{-rT}F_1\Phi(\frac{-\delta_1}{\sqrt{1+\kappa^2-2\rho \kappa}})-e^{-rT}F_2\Phi(\frac{-\delta_2}{\sqrt{1+\kappa^2-2\rho \kappa}})\\
 &-e^{-rT}K\Phi(\frac{-\delta_3}{\sqrt{1+\kappa^2-2\rho \kappa}}), \\ 
 \end{split}
\end{equation}
with $\kappa =b\frac{\sigma_2}{\sigma_1}$ and $\delta_i=z_0^i(b\sigma_2, a, x_0^i)-b\frac{\sigma_2}{\sigma_1}x_0^i, i=1,2,3$, where $z_0^i$ are given as in (\ref{barys}) and $x_0^i$ are any points.
\end{proposition}
We remark that the above relation (\ref{BS11}) is obtained by using (\ref{approx2}) and the line equations given as in (\ref{approx33}) with the points $(x_0^i, y_0^i)$  given by $(x_0^i, z_0^i(b\frac{\sigma_2}{\sigma_1}, a))$ and these latter points do not have to lie on the curves $\mathcal{C}_i$ for each $i=1,2,3$.  The proof of this proposition will be given in the appendix. It is also worthy to note that in the above proposition the choices of $x_0^i$ are irrelevant. It will be shown in the proof that, this is due to the choice of the approximations (\ref{barys}).  

\subsection{Extension of the Bjerksund-Stensland formula}
From  the relations (\ref{approx1}), (\ref{approx2}), and (\ref{approx33}) above we see that appropriate choices of the slopes $\kappa_i$ and the points $(x_0^i, y_0^i)$ on the corresponding curves $\mathcal{C}_1, \mathcal{C}_2, \mathcal{C}_3$ are important for the spread option pricing formula. From the above subsection, we see that  slopes of the form $\kappa_i=b\frac{\sigma_2}{\sigma_1}$ and the points of the form $(x_0^i, z_0^i(q, a))$ give us simple formula for the spread prices. In this section, we elaborate on the choice of the parameter $b$ in our Proposition \ref{BS-formula} above further. Our analysis will enable us to give a closed-form formula for the spread option price which is more general than the Bjerksund and Stensland spread option formula. 

To this end, consider points $(x_{\ell}^i, y_{\ell}^i)$ on the curves $\mathcal{C}_i, i=1,2,3,$ respectively. By solving for $y_{\ell}^i$ for each $i=1,2,3$ we obtain
\begin{equation}
\begin{split}
y_{\ell}^1&=\frac{1}{\sigma_1 \sqrt{T}}\ln \left(\frac{\alpha \bar{F}_2e^{\sigma_2\sqrt{T}x_{\ell}^1}+K}{g_1\bar{F}_1}\right),\\ y_{\ell}^2&=\frac{1}{\sigma_1 \sqrt{T}}\ln \left(\frac{g_2 \bar{F}_2e^{\sigma_2\sqrt{T}x_{\ell}^2}+K}{\alpha \bar{F}_1}\right),\\ y_{\ell}^3&=\frac{1}{\sigma_1 \sqrt{T}}\ln\left( \frac{\bar{F}_2e^{\sigma_2\sqrt{T}x_{\ell}^3}+K}{\bar{F}_1}\right).    
\end{split}
\end{equation}
From Lemma (\ref{lem2}) and Hopital's rule we obtain
\begin{equation}\label{bsss}
 \begin{split}
  \lim_{x_{\ell}^1\rightarrow \infty}\frac{y_{\ell}^1}{x_{\ell}^1}&=\lim_{x_{\ell}^1\rightarrow \infty}(y_{\ell}^1)'=\frac{\sigma_2}{\sigma_1}  \lim_{x_{\ell}^1\rightarrow \infty}\frac{\alpha \bar{F}_2 e^{\sigma_2\sqrt{T}x_{\ell}^1}}{\alpha \bar{F}_2e^{\sigma_2\sqrt{T}x_{\ell}^1}+K},\\
   \lim_{x_{\ell}^2\rightarrow \infty}\frac{y_{\ell}^2}{x_{\ell}^2}&=\lim_{x_{\ell}^2\rightarrow \infty}(y_{\ell}^2)'=\frac{\sigma_2}{\sigma_1}  \lim_{x_{\ell}^2\rightarrow \infty}\frac{g_2\bar{F}_2 e^{\sigma_2\sqrt{T}x_{\ell}^2}}{g_2 \bar{F}_2e^{\sigma_2\sqrt{T}x_{\ell}^2}+K},\\
    \lim_{x_{\ell}^3\rightarrow \infty}\frac{y_{\ell}^3}{x_{\ell}^3}&=\lim_{x_{\ell}^3\rightarrow \infty}(y_{\ell}^3)'=\frac{\sigma_2}{\sigma_1}  \lim_{x_{\ell}^3\rightarrow \infty}\frac{\bar{F}_2 e^{\sigma_2\sqrt{T}x_{\ell}^3}}{ \bar{F}_2e^{\sigma_2\sqrt{T}x_{\ell}^3}+K}.\\
 \end{split}   
\end{equation}
Inspired by these relations, We define the following three functions
\begin{equation}\label{bbb}
 b_1(x)=\frac{\alpha \bar{F}_2 e^{\sigma_2\sqrt{T}x}}{\alpha \bar{F}_2e^{\sigma_2\sqrt{T}x}+K}, \; \; b_2(x)=\frac{g_2\bar{F}_2 e^{\sigma_2\sqrt{T}x}}{g_2 \bar{F}_2e^{\sigma_2\sqrt{T}x}+K}, \; \; b_3(x)=\frac{\bar{F}_2e^{\sigma_2\sqrt{T}x}}{\bar{F}_2e^{\sigma_2\sqrt{T}x}+K}. 
\end{equation}
We observe that, these functions are strictly increasing on the real line and takes values in the interval $(0, 1)$.

We remark that from (\ref{bsss}) we see that the slope of the asymptotic line of the curve $\mathcal{C}_i$ is given by the limit of $\frac{\sigma_2}{\sigma_1}b_i(x)$ when $x\rightarrow +\infty$ for each $i=1,2,3$. We would like to replace the parameter $b$ in  Proposition 5.2 with the above $b_1, b_2, b_3$ in (\ref{bbb}) and obtain a more general closed-form spread option price formula. We also have to specify $a_1, a_2, a_3$. We impose the constraints $a_1(x)b_1(x)=F_2, a_2(x)b_2(x)=F_2,$ and $a_3(x)b_3(x)=F_2$, and obtain 
\begin{equation}\label{asss}
\begin{split}
a_1(x)&=F_2+Ke^{-\sigma_2 \sqrt{T}x-\rho \sigma_1 \sigma_2 T+\frac{1}{2}\sigma_2^2T},  a_2(x)=F_2+Ke^{-\sigma_2 \sqrt{T}x-\frac{1}{2}\sigma_2^2T},\\
a_3(x)&=F_2+Ke^{-\sigma_2 \sqrt{T}x+\frac{1}{2}\sigma_2^2T}.   
\end{split}
\end{equation}

Now, we are ready to state the main result of this paper. 

\begin{proposition} \label{mainmain}  When $K>0$ the price of the spread option can be approximated by
\begin{equation}\label{49}
\begin{split}
\Pi_T(&\lambda,\mu,\gamma; \theta)=:e^{-rT}F_{1}\Phi\big (\frac{-\delta_1(\lambda)}{\sqrt{1+\kappa_1^2(\lambda)-2\rho \kappa_1(\lambda)}}\big)\\
&-e^{-rT}F_{2}\Phi\big (\frac{-\delta_2(\mu)}{\sqrt{1+\kappa_2^2(\mu)-2\rho \kappa_2(\mu)}}\big)
-e^{-rT}K\Phi\big (\frac{-\delta_3(\gamma)}{\sqrt{1+\kappa_3^2(\gamma))-2\rho \kappa_3(\gamma)}}\big),\\
\end{split}
\end{equation}
for some appropriately chosen parameters $\lambda, \mu, \gamma$, where
\begin{equation}
\kappa_1(\lambda)=\frac{\sigma_2}{\sigma_1}b_1(\lambda),\; \kappa_2(\mu)=\frac{\sigma_2}{\sigma_1}b_2(\mu),\; \kappa_3(\gamma)=\frac{\sigma_2}{\sigma_1}b_3(\gamma),
\end{equation}
and 
\begin{equation}
\begin{split}
\delta_1(\lambda)&=z_0^1(\sigma_2b_1(\lambda), a_1(\lambda), \lambda)-\kappa_1(\lambda)\lambda,\\ \delta_2(\mu)&=z_0^2(\sigma_2b_2(\mu), a_2(\mu), \mu)-\kappa_2(\mu)\mu,\\ \delta_3(\gamma)&=z_0^3(\sigma_2b_3(\gamma), a_3(\gamma), \gamma)-\kappa_3(\gamma)\gamma,
\end{split}
\end{equation}
and the functions $z_0^i(\cdot), i=1,2,3,$ are given as in (\ref{barys}), $b_i(\cdot), i=1,2,3,$ are given as in (\ref{bbb}), and $\theta$ is given as in (\ref{theta}). When $K<0$, the price of the spread option can be approximated by
\begin{equation}\label{PT}
\bar{\Pi}_T(\lambda, \mu, \gamma; \theta)=:e^{-rT}(F_1-F_2-K)+\Pi_T(\lambda, \mu, \gamma; \bar{\theta}),
\end{equation}
where $\bar{\theta}$ is given by (\ref{theta}).
\end{proposition}
We remark that our closed-form formula in the above proposition depends on the three parameters $\lambda, \mu,$ and $\gamma$.  If we choose the following parameter value
\begin{equation}
\begin{split}
 \lambda&=(\frac{1}{2}\sigma_2-\rho \sigma_1)\sqrt{T}, \\
 \mu&=-\frac{1}{2}\sigma_2\sqrt{T},\\
 \gamma&= \frac{1}{2}\sigma_2\sqrt{T},
 \end{split}
\end{equation}
our formula in Proposition \ref{mainmain} reduces to the \cite{bjerksund2014closed} formula. We state this fact as a Corollary below

\begin{corollary} \label{cor52} When $K>0$, we have 
\begin{equation}
 \Pi^{BS}_T=\Pi_T(\frac{1}{2}\sigma_2\sqrt{T}-\rho \sigma_1\sqrt{T}, -\frac{1}{2}\sigma_2\sqrt{T}, \frac{1}{2}\sigma_2\sqrt{T}; \theta),  
\end{equation}
when $K<0$ we have 
\begin{equation}
 \Pi^{BS}_T=e^{-rT}(F_1-F_2-K)+\bar{\Pi}_T(\frac{1}{2}\sigma_1\sqrt{T}-\rho \sigma_2\sqrt{T}, -\frac{1}{2}\sigma_1\sqrt{T}, \frac{1}{2}\sigma_1\sqrt{T}; \theta).  
\end{equation}
\end{corollary}

\begin{remark}\label{5666} Clearly our formula (\ref{49}) generalizes the Bjerksund-Stensland formula  as explained in the Corollary \ref{cor52} above. The main challenge in applying the formula (\ref{49}) in practice is to choose optimal parameter values $\lambda, \mu, \gamma$ for the high accuracy of our approximation (\ref{49}) to the true spread prices. The optimal values of $\lambda, \mu, \gamma$ may depend on the range of model parameters. For example, high volatility $\sigma_1$ and $\sigma_2$ in both of the prices of the spread may need an optimal  $\lambda, \mu, \gamma$ while low volatility may need another optimal set of parameter values $\lambda, \mu, \gamma$. 
\end{remark}

\section{Numerical results}
In this section we test the performances of our Propositions \ref{prop2} and \ref{mainmain}. Namely, we use the Monte-Carlo (MC) method as a benchmark and compare the performance of the Proposition \ref{prop2} with the MC method and the performance of the Proposition \ref{mainmain} with both  the Bjerksund-Stensland formula and the MC method. The numerical computations for all these were implemented in Python 3.7.7 on a Legion desktop PC with 3.00GHz AMD Ryzen 5 4600H CPU.

We use the quasi-Monte Carlo method to obtain estimates for the price of the spread option in the numerical results. We use the two-dimensional Halton sequence in simulation. We use the control variate method suggested in the footnote of \cite{bjerksund2014closed}.

First, we discuss the performance of the Proposition \ref{prop2}. More specifically, we check the performance of (\ref{266S}). As stated in the Proposition \ref{prop2},  (\ref{266S}) is the application of the Simpson's rule to the Riemann integral $\int_{-\infty}^{+\infty}f(a)da$, where $f(a)$ is given as in the proposition. The error bound is given as in (\ref{SM}). Here $M$ is the maximum value of 
$f^{(4)}(a)$ on $[-b, b]$. Clearly, the value of $M$ depends on the underlying parameters of the spread option. As such the performance of $\Pi_T(N, b)$ depends on $N$ and $b$. However, our extensive numerical tests show that the value of $\Pi_T(500, 5)$ (which means $N=500$ and $b=5$)
approximates the spread price pretty well for most of the model parameters. 

In the following Table 1 we take the following model parameters $r=0.05, T=1, F_1 = 110e^{(r-0.03)T},$ $F_2 = 100e^{(r-0.02)T},\sigma_1 = 0.1, \sigma_2=0.15,$ and test our formula (\ref{266S}) in Proposition \ref{prop2} against the MC method with $\Pi_T(500, 5)$.

\begin{table}
\centering
\caption{MC vs Proposition \ref{prop2}}
\begin{threeparttable}
\centering
\begin{tabular}{l|l|l|l|l|l|l}
\hline
\diagbox{$\emph{K}$}{$\rho$}& -0.95 & -0.5 & -0.1 & 0.3 & 0.8 & 0.95 \\ \hline
-20      & 29.589165 & 28.994817 & 28.496876 & 28.070104 & 27.770086 & 27.754400 \\ 
         & \textit{29.589155} & \textit{28.994802} & \textit{28.496847} & \textit{28.070112} & \textit{27.770084} & \textit{27.754391} \\ 
         & (0.000001) & (0.000007) & (0.000007) & (0.000011) & (0.000006) & (0.000009) \\ \hline
-10      & 21.774860 & 20.904959 & 20.095212 & 19.270085 & 18.381078 & 18.256529 \\ 
         & \textit{21.774855} & \textit{20.904951} & \textit{20.095207} & \textit{19.270085} & \textit{18.381078} & \textit{18.256534} \\ 
         & (0.000001) & (0.000002) & (0.000004) & (0.000005) & (0.000006) & (0.000012) \\ \hline
-5       & 18.245477 & 17.248448 & 16.282814 & 15.230136 & 13.855745 & 13.551599 \\ 
         & \textit{18.245473} & \textit{17.248444} & \textit{16.282809} & \textit{15.230137} & \textit{13.855740} & \textit{13.551599} \\ 
         & (0.000000) & (0.000001) & (0.000001) & (0.000003) & (0.000001) & (0.000004) \\ \hline
5        & 12.122831 & 10.956214 & 9.771294 & 8.367403 & 5.967035 & 4.895080 \\ 
         & \textit{12.122832} & \textit{10.956212} & \textit{9.771296}  & \textit{8.367407}  & \textit{5.967035}  & \textit{4.895077}  \\ 
         & (0.000003) & (0.000002) & (0.000002) & (0.000003) & (0.000005) & (0.000005) \\ \hline
15       & 7.401225 & 6.242213 & 5.067175 & 3.679804 & 1.342507 & 0.366306 \\ 
         & \textit{7.401220}  & \textit{6.242210}  & \textit{5.067171}  & \textit{3.679779}  & \textit{1.342501}  & \textit{0.366314}  \\ 
         & (0.000010) & (0.000010) & (0.000011) & (0.000015) & (0.000030) & (0.000041) \\ \hline
25       & 4.098248 & 3.130020 & 2.203558 & 1.220011 & 0.104117 & 0.000326 \\ 
         & \textit{4.098247}  & \textit{3.130016}  & \textit{2.203546}  & \textit{1.220039}  & \textit{0.104096}  & \textit{0.000347}  \\ 
         & (0.000009) & (0.000016) & (0.000028) & (0.000051) & (0.000073) & (0.000047) \\ \hline
\end{tabular}
\begin{tablenotes}    
        \footnotesize 
        \item Parameters are $r=0.05,T=1, F_1 = 110e^{(r-0.03)T}, F_2 = 100e^{(r-0.02)T},\sigma_1 = 0.1, \sigma_2=0.15$.
        \item The first row is  Proposition \ref{prop2}. For the whole table it spent 0:00:00.453559. The  RMSE for the table is $1.0681e-05$.
        \item The Second row is MC simulation with $100,000$ paths. For the whole table it spent 0:00:04.313037.
        \item The Third row is the Standard Error of the MC simulation method.
\end{tablenotes}
\end{threeparttable}
\end{table}

It can be seen from Table 1 above that the formula (\ref{266S}) in  Proposition \ref{prop2} with $e^{-rT}\bar{\Pi}_T(500, 5)$ produces highly accurate spread option prices as they are very close to the spread prices obtained by the MC simulation. Also, the speed of this method is faster than the MC simulation. The time consumption for the whole table is recorded for both of these two methods, and the MC took 0:00:04:313037 seconds while the formula (\ref{266S})  in Proposition \ref{prop2} took only 0:00:00.453559 seconds.

Next, we discuss the numerical performance of the Proposition \ref{mainmain}. As stated in the Remark \ref{5666}, the performance of the formula (\ref{49}) depends on the choice of the three parameters $\lambda, \mu,$ and $\gamma$. From Corollary \ref{cor52}, if we take $\lambda_0=:\frac{1}{2}\sigma_2\sqrt{T}-\rho \sigma_1 \sqrt{T}, \mu_0=:-\frac{1}{2}\sigma_2\sqrt{T}, \gamma_0=:\frac{1}{2}\sigma_2\sqrt{T},$ then $\Pi_T(\lambda_0, \mu_0, \gamma_0; \theta)=\Pi_T^{BS}(\theta)$ for all the model parameters $\theta$. Since $\Pi_T^{BS}\le \Pi^{MC}_T$ (Here $\Pi^{MC}_T$ refers to spread prices obtained by Monte-Carlo methods), if we can find a $(\bar{\lambda},\bar{\mu}, \bar{\gamma})$ such that $\Pi_T^{BS}\le \Pi_T(\bar{\lambda}, \bar{\mu}, \bar{\gamma})\le \Pi^{MC}$ for all model parameters, then $\Pi_T(\bar{\lambda}, \bar{\mu}, \bar{\gamma})$ will be a more precise pricing formula for spread options compared to $\Pi_T^{BS}$. The determination of such $(\bar{\lambda},\bar{\mu}, \bar{\gamma})$ is a  challenging issue as the process involves six model parameters $\theta=(F_1, \sigma_1, F_2, \sigma_2, K, \rho)$. We leave further investigation of our formula in Proposition \ref{mainmain} for future work. In our discussions below, we construct a formula by using the formula in Proposition \ref{mainmain} and demonstrate that it performs better than $\Pi_T^{BS}$ for a certain range of model parameters.

To this end, for any $\theta$ given by (\ref{theta}) we define
\begin{equation}
g(x)=:\Pi_T(\frac{1}{2}\sigma_2\sqrt{T}-x \sigma_1\sqrt{T}, -\frac{1}{2}\sigma_2\sqrt{T}, \frac{1}{2}\sigma_2\sqrt{T}; \theta) ,  
\end{equation}
when $K>0$ and 
\[
g(x)=e^{-rT}(F_1-F_2-K)+ \Pi_T(\frac{1}{2}\sigma_1\sqrt{T}-x \sigma_2\sqrt{T}, -\frac{1}{2}\sigma_1\sqrt{T}, \frac{1}{2}\sigma_1\sqrt{T}; \theta),
\]
when $K<0$. Due to Corollary \ref{cor52}, we have $g(\rho)=\Pi_T^{BS}$. In the following discussions, we denote $\Pi^0_T=:g(0)$.

Our aim here is to find a proper value of $x=x_0$ such that $\Pi_T^{BS}=g(\rho)\le g(x_0)\le \Pi^{MC}_T$ for most of the model parameters. The equation of the line that passes through the two points $(0, g(0))$ and $(\rho, g(\rho))$ (here we assume $\rho \neq 0$) is given by $y-g(\rho)=\frac{g(\rho)-g(0)}{\rho}(x-\rho)$. By letting $x=\tau \rho, \tau>0, $ we obtain $y=g(\rho)+(g(\rho)-g(0))(\tau-1)$. We would like to choose $\tau$ in a proper way so that the corresponding value $y$ gives a good approximation for the spread price. Based on these analysis, we denote $\delta=:|\tau-1|$ and we introduce the following formula
\begin{equation}\label{new63}
\Pi_T^{new}=:\Pi_T^{BS}+|\Pi_T^{BS}-\Pi_T^0|\delta.   
\end{equation}
Note here that, by using the function $g(x)$, the equation (\ref{new63}) is written as $\Pi_T^{new}=g(\rho)+|g(\rho)-g(0)|\delta$. 

Clearly $\Pi_T^{BS}\le \Pi_T^{new}$ for all model parameters. To obtain a good approximation for the spread price, the $\delta$ in (\ref{new63}) needs to be selected properly so that $\Pi_T^{new}\le \Pi_T^{MC}$ for a large range of model parameters. A proper choice of $\delta$ is not an easy task as it involves six model parameters $F_1, \sigma_1, F_2, \sigma_2, K, \rho$ as mentioned earlier. However, based on extensive numerical tests we take $\delta$ as
\begin{equation}\label{delta}
\delta=\frac{T^2}{4}\frac{|K|\sigma_1^2\sigma_2^2}{F_1+F_2+|K|}(1+\rho). \end{equation}

The following Table 2 reports the performance of 
(\ref{new63}) with $\delta$ given by (\ref{delta}). In this table, MC simulations is used as a benchmark. It compares the performance of the Bjerksund-Stensland formula with the formula in (\ref{new63}).

\begin{table}
\caption{ Equation (\ref{new63}) vs BS}
\centering
\label{tab:my-table}
\begin{threeparttable}
\begin{tabular}{c|c|c|c|c|c|c}
\hline
\diagbox{$K$}{$\rho$} & -0.95 & -0.5 & -0.1 & 0.3 & 0.8 & 0.95 \\ \hline
-20 & 72.447903 & 66.789701 & 60.467212 & 52.312289 & 36.540138 & 28.179235 \\
    & 72.443623 & 66.769020 & 60.460059 & 52.280630 & 36.384937 & 27.922717 \\
    & \textit{72.452464} & \textit{66.801532} & \textit{60.525121} & \textit{52.393443} & \textit{36.580742} & \textit{28.104203} \\
    & (0.000429) & (0.001113) & (0.001847) & (0.002868) & (0.008091) & (0.010341) \\ \hline
-10 & 67.093565  & 61.310225  & 54.852982  & 46.444229  & 29.736416  & 20.177831  \\
    & 67.092422  & 61.304703  & 54.851068  & 46.435707  & 29.692196  & 20.087154  \\
    & \textit{67.096049} & \textit{61.314047} & \textit{54.869808} & \textit{46.471202} & \textit{29.773190} & \textit{20.191560}\\
    & (0.000211) & (0.000431) & (0.000679) & (0.001232) & (0.002493) & (0.005674) \\ \hline
-5  & 64.536953  & 58.716686  & 52.218566  & 43.732412  & 26.714252  & 16.619350  \\
    & 64.536659  & 58.715262  & 52.218072  & 43.730210  & 26.702613  & 16.593433  \\
    & \textit{64.537906} & \textit{58.717813} & \textit{52.222978} & \textit{43.739778} & \textit{26.728705} & \textit{16.637189}\\
    & (0.000092) & (0.000162) & (0.000249) & (0.000393) & (0.001105) & (0.001943) \\ \hline
5   & 59.662399  & 53.825895  & 47.305453  & 38.779852  & 21.592906  & 11.151212         \\
    & 59.662096  & 53.824437  & 47.304949  & 38.777608  & 21.580957  & 11.122946         \\
    & \textit{59.663653} & \textit{53.827143} & \textit{47.309946} & \textit{38.787299} & \textit{21.607722} & \textit{11.177546}\\
    & (0.000106) & (0.000169) & (0.000251) & (0.000572) & (0.000999) & (0.002857) \\ \hline
10  & 57.344219  & 51.529121  & 45.027928  & 36.541698  & 19.508779  & 9.340300          \\
    & 57.343040  & 51.523442  & 45.025965  & 36.532959  & 19.462457  & 9.233733          \\
    & \textit{57.348038} & \textit{51.533277} & \textit{45.044381} & \textit{36.567951} & \textit{19.553497} & \textit{9.366418 }\\
    & (0.000254) & (0.000443) & (0.000688) & (0.001142) & (0.003330) & (0.004886) \\ \hline
25  & 50.874145  & 45.228206  & 38.893744  & 30.730348  & 14.856800  & 6.131335          \\
    & 50.867437  & 45.195819  & 38.882554  & 30.680800  & 14.609299  & 5.675327          \\
    & \textit{50.882510} & \textit{45.242931} & \textit{38.975629} & \textit{30.838972} & \textit{14.867635} & \textit{5.961783 }\\
    & (0.000626) & (0.001467) & (0.002414) & (0.003837) & (0.008857) & (0.014012) \\ \hline
\end{tabular}
\begin{tablenotes}   
        \footnotesize          \item Parameters are $r=0.0125,T=4, F_1 = 110e^{(r-0.03)T}, F_2 = 100e^{(r-0.02)T},\sigma_1 = 0.45, \sigma_2=0.45$.     
        \item   The first row corresponds to Equation (\ref{new63}) (the RMSE for the whole table is 0.04499244). 
        \item The second row corresponds to Bjerksund-Stensland formula (the RMSE for the whole table is 0.09539263). 
        \item The third row (in italics) is the MC simulation (100,000 trials). 
        \item The fourth row (in parentheses) is the standard error of the simulation. \end{tablenotes}      
\end{threeparttable}
\end{table}

As it can be seen from the above Table 2, the RMSE for the Bjerksund-Stensland formula is 0.09539263 while the RMSE for our formula (\ref{new63}) is 0.04499244 for the whole table.
Here, in the calculations of RMSE, we used MC prices ( with 100,000 trials) as a benchmark, i.e.,
\begin{equation}\label{RMSE}
 RMSE=\sqrt{\frac{\sum_i^n(p_i-MC_i)^2}{n} },   
\end{equation}
where $p_i$ are the corresponding prices (the prices obtained by (\ref{new63}) and by the Bjerksund-Stensland  formula separately) and $n$ is the number of pairs of $\rho$ and $K$ (in the above table $n=36$). Also, it can be seen from the above Table 2 that our formula (\ref{new63}) improves the Bjerksund-Stensland formula for each single case.

In the following Table 3, we compare the RMSEs of (\ref{new63}) and the Bjerksund-Stensland formula for different pairs of volatility. These are RMSE values for the whole of Table 2 (keeping all the other parameters the same as in Table 2) for different pairs of volatility $\sigma_1$ and $\sigma_2$. To calculate these RMSE values, we fixed some of the parameters at $r=0.0125,T=4, F_1 = 110e^{(r-0.03)T}, F_2 = 100e^{(r-0.02)T}$  as in Table 2 above first and for each pair $(\sigma_1, \sigma_2)$ in Table 3  we calculated the corresponding prices for the model parameters $\theta(i, j)=:(F_1, \sigma_1, F_2, \sigma_2, K_i, \rho_j)$, $1\le i\le 6$, $1\le j\le 6$, where $K_i$  covers the strikes in Table 2 above and $\rho_j$ covers the values of the correlations in Table 2. This gives us a total of 36 prices for each pair of volatility. Then we applied (\ref{RMSE}), using the 
 MC prices (with 100,000 trials) as a benchmark.
 
\begin{table}\label{TTTO}
\caption{Comparisons of RMSEs}
\centering
\begin{threeparttable}
\begin{tabular}{l|l|l|l|l|l|l|l|l}
\hline
\diagbox{$\sigma_2$}{$\sigma_1$} & 0.2 & 0.3 & 0.4 & 0.5 & 0.6 & 0.7 & 0.8 & 0.9  \\ \hline
0.2 & 0.005909 & 0.029058 & 0.030207 & 0.046945 & 0.081544 & 0.128063 & 0.182532 & 0.244051 \\
    & 0.007179 & 0.034590 & 0.039178 & 0.059074 & 0.098215 & 0.150129 & 0.210959 & 0.279889 \\ \hline
0.3 & 0.020722 & 0.016167 & 0.046015 & 0.092183 & 0.092402 & 0.113108 & 0.150941 & 0.196935 \\
    & 0.023713 & 0.027919 & 0.079596 & 0.145773 & 0.157443 & 0.192350 & 0.248998 & 0.317598 \\ \hline
0.4 & 0.017899 & 0.039630 & 0.027755 & 0.056536 & 0.129770 & 0.163619 & 0.172579 & 0.190030 \\
    & 0.022821 & 0.059646 & 0.067248 & 0.149603 & 0.296980 & 0.367740 & 0.410464 & 0.468561 \\ \hline
0.5 & 0.028926 & 0.060740 & 0.059701 & 0.078288 & 0.124226 & 0.124781 & 0.184020 & 0.211320 \\
    & 0.035954 & 0.091916 & 0.116627 & 0.129775 & 0.246966 & 0.462468 & 0.636327 & 0.733526 \\ \hline
0.6 & 0.051932 & 0.056810 & 0.113127 & 0.121517 & 0.208418 & 0.297181 & 0.269992 & 0.232497 \\
    & 0.062087 & 0.092920 & 0.211088 & 0.200322 & 0.217517 & 0.370571 & 0.644388 & 0.923345 \\ \hline
0.7 & 0.082681 & 0.071687 & 0.118770 & 0.158984 & 0.269682 & 0.432240 & 0.591094 & 0.637889 \\
    & 0.096480 & 0.117257 & 0.235437 & 0.357144 & 0.312307 & 0.333458 & 0.518622 & 0.834001 \\ \hline
0.8 & 0.119084 & 0.097093 & 0.121044 & 0.188299 & 0.269457 & 0.513058 & 0.765237 & 0.986953 \\
    & 0.137104 & 0.156257 & 0.255172 & 0.446486 & 0.522409 & 0.449944 & 0.471600 & 0.705054 \\ \hline
0.9 & 0.160631 & 0.128040 & 0.130843 & 0.191989 & 0.270768 & 0.506890 & 0.879435 & 1.210983 \\
    & 0.183463 & 0.202490 & 0.291729 & 0.488869 & 0.694849 & 0.702709 & 0.603389 & 0.633085\\ \hline
\end{tabular}
\begin{tablenotes}    
        \footnotesize    
       \item The first row corresponds to RMSEs of  (\ref{new63}) (RMSE for the whole Table 2).
        \item The second row corresponds to RMSEs of the Bjerksund-Stensland formula (RMSE for the whole of Table 2). 
        \item Benchmark is MC simulation (100,000 trials). 
      \end{tablenotes}           
\end{threeparttable}
\end{table}

Table 3 above shows that our formula (\ref{new63}) consistently performs better than the Bjerksund-Stensland formula under the RMSE criteria for most of the cases. 

We remark here that we do not claim that this particular choice of $\delta$ in (\ref{delta}) above is optimal for obtaining an approximate spread option valuation formula. The relation (\ref{new63}) suggests that the $\delta$ needs to be chosen as
\begin{equation}\label{deltatheta}
\frac{\Pi_T^{MC}-\Pi_T^{BS}}{|\Pi_T^{BS}-\Pi_T^0|}=\delta(F_1, \sigma_1, F_2, \sigma_2, K, \rho),   \end{equation}
(here we assume $\rho\neq 0$) for all the model parameters to match the MC values through (\ref{new63}). Obtaining an expression for $\delta$ that satisfies (\ref{deltatheta}) for all the model parameters is a difficult issue. Our choice (\ref{delta}) is based on observing the values of the left-hand-side of the expression (\ref{deltatheta}) in our numerical tests. However, as mentioned earlier, the choice of $\delta$ in (\ref{delta}) is not optimal, and we leave further discussions of these 
issues for future work. Here, we only attempt to demonstrate that the formula (\ref{49})
can be used to improve the accuracy of the Bjerksund-Stensland formula.

\section{Conclusion}
In this paper, we give an alternative derivation for the Bjerksund-Stensland formula for log-normal models. Our approach is based on the approximation of the exercise boundary of the spread by linear functions whose slopes are chosen to be equal to the slope of the asymptotic lines of the exercise boundaries after some adjustment and that, in the meantime, pass through properly selected points 
to achieve high accuracy in the approximation. By doing this, we also developed a closed-form formula that contains the Bjerksund-Stensland formula as a special case, as explained in the Corollary \ref{cor52}. Unlike the Bjerksund-Stensland and Carmona-Durrleman formulas,  our formula in Proposition \ref{mainmain} does not produce a lower or upper bound for spread price, making it challenging to choose the optimal parameter values  $\lambda, \mu, \gamma$ for high accuracy of the approximation. It is expected that such optimal values of $\lambda, \mu, \gamma$  depend on the range of model parameters of the spread. In our numerical analysis section, we give an approximation formula for spread option price as an application of our formula  (\ref{49}) and show that it performs better than the Bjerksund-Stensland formula for some range of model parameters.

\vspace{0.1in}

\textbf{\Large{Appendix A:} The Kirk formula}

\vspace{0.1in}
In this appendix we review an approximation that was used to derive Kirk's formula as we use a similar idea of this approximation in (\ref{yapprox}).

In Kirk's approximation the sum $S_2(T)+K$ of the second asset price $S_2(T)$ with the constant strike $K$ is approximated by a log-normal random variable. This approach reduces the spread price with strike $K$ into an exchange option and hence the spread price can be approximated by Margrabe's formula \cite{margrabe1978value}. Namely, if one implements the following approximation 

\begin{equation}\label{approx}
F_2e^{-\frac{1}{2}\sigma_2^2T+\sigma_2\sqrt{T}X}+K\approx (F_2+K)e^{-\frac{1}{2}\theta^2T+\theta \sqrt{T} X},
\end{equation}
with $\theta=\frac{F_2}{F_2+K}\sigma_2$ to the spread price in (\ref{6}) and then apply the Margrabe's formula one can obtain Kirk's following formula
\begin{equation*}
\begin{split}
\Pi_T^K&=E\big [ F_1e^{-\frac{1}{2}\sigma_1^2T+\sigma_1\sqrt{T}Y}-(F_2+K)e^{-\frac{1}{2}\theta^2T+\theta \sqrt{T}X} \big ]^+\\
&=e^{-rT}\big \{F_1\Phi(d_{K,1})+(F_2+K)\Phi(d_{K,2}) \big \},\\
\end{split}
\end{equation*}
where
\begin{equation}
\begin{split}
&d_{K, 1}=\frac{\ln \frac{F_1}{F_2+K}+\frac{1}{2}\sigma_K^2T}{\sigma_K\sqrt{T}},\\
&d_{K, 2}=d_{K, 1}-\sigma_K\sqrt{T},\\
&\sigma_K=\sqrt{\sigma_1^2-2\frac{F_2}{F_2+K}\rho \sigma_1 \sigma_2+\big (\frac{F_2}{F_2+K}\big )^2\sigma_2^2}.
\end{split}    
\end{equation}

Note here that the approximation (\ref{approx}) uses the weighted average $\frac{F_2}{F_2+K}\sigma_2+\frac{K}{F_2+K} 0$ of the volatility $\sigma_2$ of $\sigma_2 W_T^2$ and the volatility $0$ (which corresponds to K) for $\theta$ in (\ref{approx}). 


\vspace{0.1in}

\textbf{\Large{Appendix B:} The Greek parameters}
\vspace{0.1in}

The Greeks of option pricing formulas are important for risk management and hedging issues. In this section, we calculate the Greek parameters for our formula. First, observe that our formula (\ref{49}) can also be written as follows 
 \begin{equation}
\Pi_{T}(\lambda, \mu, \gamma)=e^{-rT}F_{1}\Phi(I)-e^{-rT}F_{2}\Phi(J)-e^{-rT}K\Phi(H),
\end{equation}
where
\begin{equation}
 \begin{split}
 I=&\frac{\ln(\frac{F_{1}}{a_{1}})+(\frac{1}{2}\sigma_{1}^{2}-\rho\sigma_{1}\sigma_{2}b_{1}+\frac{1}{2}\sigma_{2}^{2}b_{1}^{2})T}
{\bar{\sigma}_{1}\sqrt{T}},\\    
J=&\frac{\ln(\frac{F_{1}}{a_{2}})+(-\frac{1}{2}\sigma_{1}^{2}+\rho\sigma_{1}\sigma_{2}+\frac{1}{2}\sigma_{2}^{2}b_{2}^{2}-\sigma_{2}^{2}b_{2})T}
{\bar{\sigma}_{2}\sqrt{T}},\\
H=&\frac{\ln(\frac{F_{1}}{a_{3}})+(-\frac{1}{2}\sigma_{1}^{2}+\frac{1}{2}\sigma_{2}^{2}b_{3}^{2})T}
{\bar{\sigma}_{3}\sqrt{T}},
\end{split}   
\end{equation}
with
\begin{equation}
\begin{split}
\bar{\sigma}_{1}=&\sqrt{\sigma_{1}^{2}-2\rho\sigma_{1}\sigma_{2}b_{1}+\sigma_{2}^{2}b_{1}^{2}},\\
 \bar{\sigma}_{2}=&\sqrt{\sigma_{1}^{2}-2\rho\sigma_{1}\sigma_{2}b_{2}+\sigma_{2}^{2}b_{2}^{2}},\\
 \bar{\sigma}_{3}=&\sqrt{\sigma_{1}^{2}-2\rho\sigma_{1}\sigma_{2}b_{3}+\sigma_{2}^{2}b_{3}^{2}},\\
\end{split}    
\end{equation}
and the functions $b_1(\lambda), b_2(\mu), b_3(\gamma)$ are given as in (\ref{bbb}). 

Clearly, this formula is similar to the Bjerksund and Stensland formula (\ref{BSS2}) above. To find the Greeks, we assume  $b_1, b_2, b_3$ are constants as in Bjerksund and Stensland \cite{bjerksund2014closed} and $a_1, a_2, a_3$ are given as in (\ref{asss}). Under this assumptions, we calculate the Greeks as follows.

The Delta parameters of our formula are given as follows
\begin{equation}
\begin{split}
 \frac{\partial\Pi_{T}}{\partial F_{1}}=&e^{-rT}\Phi(I)+e^{-rT}\left\{\frac{F_{1}\phi(I)}{\bar{\sigma}_{1}\sqrt{T}}-\frac{F_{2}\phi(J)}{\bar{\sigma}_{2}\sqrt{T}}-\frac{K\phi(H)}{\bar{\sigma}_{3}\sqrt{T}}\right\}\frac{1}{F_{1}},\\ 
\frac{\partial\Pi_{T}}{\partial F_{2}}=&-e^{-rT}\Phi(J)-e^{-rT}\left\{\frac{F_{1}\phi(I)}{\bar{\sigma}_{1}\sqrt{T}}\frac{a_{1}'}{a_{1}}-\frac{F_{2}\phi(J)}{\bar{\sigma}_{2}\sqrt{T}}\frac{a_{2}'}{a_{2}}-\frac{K\phi(H)}{\bar{\sigma}_{3}\sqrt{T}}\frac{a_{3}'}{a_{3}}\right\}\\
=&-e^{-rT}\Phi(J)-e^{-rT}\left\{\frac{F_{1}\phi(I)}{\bar{\sigma}_{1}\sqrt{T}}\frac{b_{1}}{F_{2}}-\frac{F_{2}\phi(J)}{\bar{\sigma}_{2}\sqrt{T}}\frac{b_{2}}{F_{2}}-\frac{K\phi(H)}{\bar{\sigma}_{3}\sqrt{T}}\frac{b_{3}}{F_{2}}\right\}.
\end{split}    
\end{equation}
The Gamma parameters of our formula are given as follows
\begin{align}
\begin{split}
 \frac{\partial^{2}\Pi_{T}}{\partial F_{1}^{2}}=&2e^{-rT}\phi(I)\frac{1}{\bar{\sigma}_{1}\sqrt{T}F_{1}}\\
   -&e^{-rT}\left\{F_{1}\phi(I)I\frac{1}{\bar{\sigma}_{1}^{2}T}-F_{2}\phi(J)J\frac{1}{\bar{\sigma}_{2}^{2}T}
-K\phi(H)H\frac{1}{\bar{\sigma}_{3}^{2}T}\right\}\frac{1}{F_{1}^{2}}\\
-&e^{-rT}\left\{\frac{F_{1}\phi(I)}{\bar{\sigma}_{1}\sqrt{T}}-\frac{F_{2}\phi(J)}{\bar{\sigma}_{2}\sqrt{T}}-\frac{K\phi(H)}{\bar{\sigma}_{3}\sqrt{T}}\right\}\frac{1}{F_{1}^{2}},\\
\frac{\partial^{2}\Pi_{T}}{\partial F_{1}\partial F_{2}}=&-e^{-rT}\phi(I)\frac{a_{1}'}{a_{1}\bar{\sigma}_{1}\sqrt{T}}-e^{-rT}\phi(J)\frac{1}{F_{1}\bar{\sigma}_{2}\sqrt{T}}\\
+&e^{-rT}\left\{\frac{F_{1}\phi(I)I}{\bar{\sigma}_{1}^{2}T}\frac{a_{1}'}{a_{1}}-\frac{F_{2}\phi(J)J}{\bar{\sigma}_{2}^{2}T}\frac{a_{2}'}{a_{2}}-\frac{K\phi(H)H}{\bar{\sigma}_{3}^{2}T}\frac{a_{3}'}{a_{3}}\right\}\frac{1}{F_{1}}\\
=&-e^{-rT}\phi(I)\frac{b_{1}}{F_{2}\bar{\sigma}_{1}\sqrt{T}}-e^{-rT}\phi(J)\frac{1}{F_{1}\bar{\sigma}_{2}\sqrt{T}}\\
+&e^{-rT}\left\{\frac{F_{1}\phi(I)I}{\bar{\sigma}_{1}^{2}T}\frac{b_{1}}{F_{2}}-\frac{F_{2}\phi(J)J}{\bar{\sigma}_{2}^{2}T}\frac{b_{2}}{F_{2}}-\frac{K\phi(H)H}{\bar{\sigma}_{3}^{2}T}\frac{b_{3}}{F_{2}}\right\}\frac{1}{F_{1}},\\
\end{split}    
\end{align}
\begin{align}
\begin{split}
\frac{\partial^{2}\Pi_{T}}{\partial F_{2}^{2}}=&2e^{-rT}\frac{\phi(J)}{\bar{\sigma}_{2}\sqrt{T}}\frac{a_{2}'}{a_{2}}\\
-e^{-rT}&\left\{\frac{F_{1}\phi(I)I}{\bar{\sigma}_{1}^{2}T}(\frac{a_{1}'}{a_{1}})^{2}
-\frac{F_{2}\phi(J)J}{\bar{\sigma}_{2}^{2}T}(\frac{a_{2}'}{a_{2}})^{2}-\frac{K\phi(H)H}{\bar{\sigma}_{3}^{2}T}(\frac{a_{3}'}{a_{3}})^{2}\right\}\\
-e^{-rT}&\left\{\frac{F_{1}\phi(I)}{\bar{\sigma}_{1}\sqrt{T}}\frac{a_{1}''a_{1}-(a_{1}')^{2}}{a_{1}^{2}}
-\frac{F_{2}\phi(J)}{\bar{\sigma}_{2}\sqrt{T}}\frac{a_{2}''a_{2}-(a_{2}')^{2}}{a_{2}^{2}}
-\frac{K\phi(H)}{\bar{\sigma}_{3}\sqrt{T}}\frac{a_{3}''a_{3}-(a_{3}')^{2}}{a_{3}^{2}}\right\}\\
=2e^{-rT}&\frac{\phi(J)}{\bar{\sigma}_{2}\sqrt{T}}\frac{b_{2}}{F_{2}}-e^{-rT}\left\{\frac{F_{1}\phi(I)I}{\bar{\sigma}_{1}^{2}T}(\frac{b_{1}}{F_{2}})^{2}
-\frac{F_{2}\phi(J)J}{\bar{\sigma}_{2}^{2}T}(\frac{b_{2}}{F_{2}})^{2}-\frac{K\phi(H)H}{\bar{\sigma}_{3}^{2}T}(\frac{b_{3}}{F_{2}})^{2}\right\}\\
+&e^{-rT}\left\{\frac{F_{1}\phi(I)}{\bar{\sigma}_{1}\sqrt{T}}(\frac{b_{1}}{F_{2}})^{2}
-\frac{F_{2}\phi(J)}{\bar{\sigma}_{2}\sqrt{T}}(\frac{b_{2}}{F_{2}})^{2}
-\frac{K\phi(H)}{\bar{\sigma}_{3}\sqrt{T}}(\frac{b_{3}}{F_{2}})^{2}\right\}.\\
\end{split}    
\end{align}

The Vegas are given by
\begin{equation}
\begin{split}
 \frac{\partial\Pi_{T}}{\partial\sigma_{1}}=&e^{-rT}\left\{F_{1}\phi(I)\frac{(\sigma_{1}-\rho\sigma_{2}b_{1})T}{\bar{\sigma}_{1}\sqrt{T}}
+F_{2}\phi(J)\frac{(\sigma_{1}-\rho\sigma_{2})T}{\bar{\sigma}_{2}\sqrt{T}}+K\phi(H)\frac{\sigma_{1}T}{\bar{\sigma}_{3}\sqrt{T}}\right\}\\   
-&e^{-rT}\left\{F_{1}\phi(I)I\frac{\sigma_{1}-\rho\sigma_{2}b_{1}}{\bar{\sigma}_{1}^{2}}-F_{2}\phi(J)J\frac{\sigma_{1}-\rho\sigma_{2}b_{2}}{\bar{\sigma}_{2}^{2}}
-K\phi(H)H\frac{\sigma_{1}-\rho\sigma_{2}b_{3}}{\bar{\sigma}_{3}^{2}}\right\}.\\
\frac{\partial \Pi_{T}}{\partial\sigma_{2}}=&\\
e^{-rT}&\left\{F_{1}\phi(I)\frac{(-\rho\sigma_{1}b_{1}+\sigma_{2}b_{1}^{2})T}{\bar{\sigma}_{1}\sqrt{T}}
-F_{2}\phi(J)\frac{(\rho\sigma_{1}+\sigma_{2}b_{2}^{2}-2\sigma_{2}b_{2})T}{\bar{\sigma}_{2}\sqrt{T}}-K\phi(H)\frac{\sigma_{2}b_{3}^{2}T}{\bar{\sigma}_{3}\sqrt{T}}\right\}-\\
e^{-rT}&\left\{F_{1}\phi(I)I\frac{-\rho\sigma_{1}b_{1}+\sigma_{2}b_{1}^{2}}{\bar{\sigma}_{1}^{2}}-F_{2}\phi(J)J\frac{-\rho\sigma_{1}b_{2}+\sigma_{2}b_{2}^{2}}{\bar{\sigma}_{2}^{2}}
-K\phi(H)H\frac{-\rho\sigma_{1}b_{3}+\sigma_{2}b_{3}^{2}}{\bar{\sigma}_{3}^{2}}\right\}.\\
\end{split}
\end{equation}

The $\Theta$ of our formula is given by 

\begin{equation}
 \begin{split}
 \frac{\partial\Pi_{T}}{\partial T}=&-r\Pi_{T}\\
 +&e^{-rT}\left\{F_{1}\phi(I)\frac{\frac{1}{2}\sigma_{1}^{2}-\rho\sigma_{1}\sigma_{2}b_{1}+\frac{1}{2}\sigma_{2}^{2}b_{1}^{2}}{\bar{\sigma}_{1}\sqrt{T}}
-F_{2}\phi(J)\frac{-\frac{1}{2}\sigma_{1}^{2}+\rho\sigma_{1}\sigma_{2}+\frac{1}{2}\sigma_{2}^{2}b_{2}^{2}-\sigma_{2}^{2}b_{2}}{\bar{\sigma}_{2}\sqrt{T}}\right.\\
&\left.-K\phi(H)\frac{-\frac{1}{2}\sigma_{1}^{2}+\frac{1}{2}\sigma_{2}^{2}b_{3}^{2}}{\bar{\sigma}_{3}\sqrt{T}}\right\}-e^{-rT}\{F_{1}\phi(I)I-F_{2}\phi(J)J-K\phi(H)H\}\frac{1}{2T}.\\
 \end{split}   
\end{equation}

The $\rho$ is given by
\begin{equation}
 \frac{\partial\Pi_{T}}{\partial r}=-T\{e^{-rT}F_{1}\Phi(I)-e^{-rT}F_{2}\Phi(J)-e^{-rT}K\Phi(H)\}=-T\Pi_{T}.    
\end{equation}

The partial derivative w.r.t $\rho$ is given by
\begin{equation}
 \begin{split}
 \frac{\partial\Pi_{T}}{\partial\rho}=&e^{-rT}\left\{-F_{1}\phi(I)\frac{b_{1}\sigma_{1}\sigma_{2}T}{\bar{\sigma}_{1}\sqrt{T}}-F_{2}\phi(J)
\frac{\sigma_{1}\sigma_{2}T}{\bar{\sigma}_{2}\sqrt{T}}\right\}\\
+&e^{-rT}\left\{\frac{F_{1}\phi(I)Ib_{1}}{\bar{\sigma_{1}^{2}}}-\frac{F_{2}\phi(J)Jb_{2}}{\bar{\sigma}_{2}^{2}}-\frac{K\phi(H)Hb_{3}}{\bar{\sigma}_{3}^{2}}\right\}\sigma_{1}\sigma_{2}.\\
 \end{split}   
\end{equation}

It can be verified that the Greeks satisfy the following partial differential equations:

\begin{equation}
 \frac{1}{2}\sigma_{1}^{2}F_{1}^{2}\frac{\partial^{2}\Pi_{T}}{\partial F_{1}^{2}}+\rho\sigma_{1}F_{1}\sigma_{2}F_{2}\frac{\partial^{2}\Pi_{T}}{\partial F_{1}\partial F_{2}}+\frac{1}{2}\sigma_{2}^{2}F_{2}^{2}\frac{\partial^{2}\Pi_{T}}{\partial F_{2}^{2}}-\frac{\partial\Pi_{T}}{\partial T}=r\Pi_{T}.   
\end{equation}
and
\begin{equation}
 \frac{1}{2}\sigma_{1}\frac{\partial\Pi_{T}}{\partial\sigma_{1}}+\frac{1}{2}\sigma_{2}\frac{\partial\Pi_{T}}{\partial\sigma_{2}}
+r\frac{\partial\Pi_{T}}{\partial r}=T\frac{\partial\Pi_{T}}{\partial T}.   \end{equation}

\noindent \textbf{\Large{Appendix C:}  Proofs}
\vspace{0.1in}

\textbf{Proof of Proposition \ref{41}:}
From (\ref{RC2}) we have 
\begin{equation}
\begin{split}
 \bar{\Pi}_{T}^{CD}(\theta_0,d_0)=F_{1}\Phi(d_0+\sigma_{1}\sqrt{T}\cos(\theta_0+\phi))-F_{2}\Phi(d_0+\sigma_{2}\sqrt{T}\cos(\theta_0))-K\Phi(d_0).
\end{split}
\end{equation}
Let $d_1, d_2, d_3,$ denote the values in  (16), (17), (18) of \cite{bjerksund2014closed} (  Bjerksund and Stensland formula). Observe that $d_0=d_3$. Also note that 
\begin{equation}
\begin{split}
 d_{0}+\sigma_{2}\sqrt{T}\cos\theta_{0}=&\frac{\ln\frac{F_{1}}{a}-\frac{1}{2}\sigma_{1}^{2}T+\frac{1}{2}b^{2}\sigma_{2}^{2}T}{\sqrt{\sigma_{1}^{2}+\sigma_{2}^{2}b^{2}-2\sigma_{1}\sigma_{2}b\cos\phi}\sqrt{T}}
 +\sigma_{2}\sqrt{T}\frac{-\sigma_{2}b+\sigma_{1}\cos\phi}{\sqrt{\sigma_{1}^{2}+\sigma_{2}^{2}b^{2}-2\sigma_{1}\sigma_{2}b\cos\phi}}\\
 =&\frac{\ln\frac{F_{1}}{a}-\frac{1}{2}\sigma_{1}^{2}T+\frac{1}{2}b^{2}\sigma_{2}^{2}T-b\sigma_{2}^{2}T+\sigma_{1}\sigma_{2}T\cos\phi}{\sqrt{\sigma_{1}^{2}+\sigma_{2}^{2}b^{2}-2\sigma_{1}\sigma_{2}b\cos\phi}\sqrt{T}}=d_2,
\end{split}
\end{equation}
and
\begin{equation}
\begin{split}
 &d_{0}+\sigma_{1}\sqrt{T}\cos(\theta_{0}+\phi)\\
 =&d_{0}+\sigma_{1}\sqrt{T}\cos\theta_{0}\cos\phi-\sigma_{1}\sqrt{T}\sin\theta_{0}\sin\phi\\
 =&\frac{\ln\frac{F_{1}}{a}-\frac{1}{2}\sigma_{1}^{2}T+\frac{1}{2}b^{2}\sigma_{2}^{2}T}{\sqrt{\sigma_{1}^{2}+\sigma_{2}^{2}b^{2}-2\sigma_{1}\sigma_{2}b\cos\phi}\sqrt{T}}
 +\sigma_{1}\sqrt{T}\frac{(-\sigma_{2}b+\sigma_{1}\cos\phi)\cos\phi}{\sqrt{\sigma_{1}^{2}+\sigma_{2}^{2}b^{2}-2\sigma_{1}\sigma_{2}b\cos\phi}}\\
 -&\sigma_{1}\sqrt{T}\frac{-\sigma_{1}\sin^{2}\phi}{\sqrt{\sigma_{1}^{2}+\sigma_{2}^{2}b^{2}-2\sigma_{1}\sigma_{2}b\cos\phi}}\\
 =&\frac{\ln\frac{F_{1}}{a}-\frac{1}{2}\sigma_{1}^{2}T+\frac{1}{2}b^{2}\sigma_{2}^{2}T-\sigma_{1}\sigma_{2}bT\cos\phi+\sigma_{1}^{2}T\cos^{2}\phi+\sigma_{1}^{2}T\sin^{2}\phi}{\sqrt{\sigma_{1}^{2}+\sigma_{2}^{2}b^{2}-2\sigma_{1}\sigma_{2}b\cos\phi}\sqrt{T}}=d_1.
\end{split}
\end{equation}
This completes the proof.

\textbf{Proof of Proposition \ref{prop1}}:
Let 
\[
A=\{F_1e^{-\frac{1}{2}\sigma^2_1T+\sigma_1\sqrt{T}Y}-F_2e^{-\frac{1}{2}\sigma^2_2T+\sigma_2\sqrt{T}X} -K>0\},
\]
then we have
\begin{equation}\label{81}
\begin{split}
&E[(F_1e^{-\frac{1}{2}\sigma^2_1T+\sigma_1\sqrt{T}Y}-F_2e^{-\frac{1}{2}\sigma^2_2T+\sigma_2\sqrt{T}X} -K)^+]\\ &=E[(F_1e^{-\frac{1}{2}\sigma^2_1T+\sigma_1\sqrt{T}Y}-F_2e^{-\frac{1}{2}\sigma^2_2T+\sigma_2\sqrt{T}X} -K)1(A)],   
\end{split}
\end{equation}
where $1(A)$ is the indicator function of $A$. The right-hand-side of (\ref{81}) can be written as
\begin{equation}\label{67}
\begin{split}
 &E[(F_1e^{-\frac{1}{2}\sigma^2_1T+\sigma_1\sqrt{T}Y}-F_2e^{-\frac{1}{2}\sigma^2_2T+\sigma_2\sqrt{T}X} -K)^+]\\
 =&F_{1}E\left[e^{-\frac{1}{2}\sigma^2_1T+\sigma_1\sqrt{T}Y}I(F_1e^{-\frac{1}{2}\sigma^2_1T+\sigma_1\sqrt{T}Y}-F_2e^{-\frac{1}{2}\sigma^2_2T+\sigma_2\sqrt{T}X} -K\geq0)\right]\\
 -&F_{2}E\left[e^{-\frac{1}{2}\sigma^2_2T+\sigma_2\sqrt{T}X}I(F_1e^{-\frac{1}{2}\sigma^2_1T+\sigma_1\sqrt{T}Y}-F_2e^{-\frac{1}{2}\sigma^2_2T+\sigma_2\sqrt{T}X} -K\geq0)\right]\\
 -&KQ\left(F_1e^{-\frac{1}{2}\sigma^2_1T+\sigma_1\sqrt{T}Y}-F_2e^{-\frac{1}{2}\sigma^2_2T+\sigma_2\sqrt{T}X} -K\geq0\right).
\end{split}
\end{equation}
By using Girsanov's change of measure theorem, the first two terms of the right-hand-side of (\ref{67}) can be calculated as follows
\begin{equation}
\begin{split}
 &F_{1}E\left[e^{-\frac{1}{2}\sigma^2_1T+\sigma_1\sqrt{T}Y}I(F_1e^{-\frac{1}{2}\sigma^2_1T+\sigma_1\sqrt{T}Y}-F_2e^{-\frac{1}{2}\sigma^2_2T+\sigma_2\sqrt{T}X} -K\geq0)\right]\\
 =&F_{1}E\left[I(F_1e^{-\frac{1}{2}\sigma^2_1T+\sigma_1\sqrt{T}(Y+\sigma_{1}\sqrt{T})}-F_2e^{-\frac{1}{2}\sigma^2_2T+\sigma_2\sqrt{T}(X+\rho\sigma_{1}\sqrt{T})} -K\geq0)\right]\\
 =&F_{1}Q\left(F_{1}e^{\frac{1}{2}\sigma_{1}^{2}T+\sigma_{1}\sqrt{T}Y}-F_{2}e^{-\frac{1}{2}\sigma_{2}^{2}T+\rho\sigma_{1}\sigma_{2}T+\sigma_{2}\sqrt{T}X}-K\geq0\right)\\
 =&F_{1}Q\left(g_{1}\bar{F}_{1}e^{\sigma_{1}\sqrt{T}Y}-\alpha\bar{F}_{2}e^{\sigma_{2}\sqrt{T}X}-K\geq0\right),
\end{split}
\end{equation}
and
\begin{equation}
\begin{split}
 &F_{2}E\left[e^{-\frac{1}{2}\sigma^2_2T+\sigma_2\sqrt{T}X}I(F_1e^{-\frac{1}{2}\sigma^2_1T+\sigma_1\sqrt{T}Y}-F_2e^{-\frac{1}{2}\sigma^2_2T+\sigma_2\sqrt{T}X} -K\geq0)\right]\\
 =&F_{2}E\left[I(F_1e^{-\frac{1}{2}\sigma^2_1T+\sigma_1\sqrt{T}(Y+\rho\sigma_{2}\sqrt{T})}-F_2e^{-\frac{1}{2}\sigma^2_2T+\sigma_2\sqrt{T}(X+\sigma_{2}\sqrt{T})} -K\geq0)\right]\\
 =&F_{2}Q\left(F_{1}e^{-\frac{1}{2}\sigma_{1}^{2}T+\rho\sigma_{1}\sigma_{2}T+\sigma_{1}\sqrt{T}Y}-F_{2}e^{\frac{1}{2}\sigma_{2}^{2}T+\sigma_{2}\sqrt{T}X}-K\geq0\right)\\
 =&F_{2}Q\left(\alpha\bar{F}_{1}e^{\sigma_{1}\sqrt{T}Y}-g_{2}\bar{F}_{2}e^{\sigma_{2}\sqrt{T}X}-K\geq0\right).
\end{split}
\end{equation}
The last term of the right-hand-side of (\ref{67}) equals to
\begin{equation}
\begin{split}
 &KQ\left(F_1e^{-\frac{1}{2}\sigma^2_1T+\sigma_1\sqrt{T}Y}-F_2e^{-\frac{1}{2}\sigma^2_2T+\sigma_2\sqrt{T}X} -K\geq0\right)\\
 =&KQ\left(\bar{F}_{1}e^{\sigma_{1}\sqrt{T}Y}-\bar{F}_{2}e^{\sigma_{2}\sqrt{T}X}-K\geq0\right).
\end{split}
\end{equation}\\ \\

\textbf{Proof of Proposition \ref{prop2}:} 
From Proposition \ref{prop1} we have (\ref{C}). Observe that $Y|X=\theta \sim N(\theta \rho, 1-\rho^2)$. We use this to 
write each $C_D^1, C_D^2, C_D^3$ as follows, 
\begin{equation*}
\begin{split}
 C_D^1=&\int_{-\infty}^{+\infty}
 Q\left(g_1\bar{F}_{1}e^{\sigma_{1}\sqrt{T}Y}-\alpha \bar{F}_{2}e^{\sigma_{2}\sqrt{T}x}-K\geq 0 \right)\varphi(x)dx\\
 =&\int_{-\infty}^{+\infty}Q\left(Y\geq\frac{\ln\left(\frac{\alpha\bar{F}_{2}}{g_{1}\bar{F}_{1}}e^{\sigma_{2}\sqrt{T}x}+\frac{K}{g_{1}\bar{F}_{1}}\right)}{\sigma_{1}\sqrt{T}}\right)\varphi(x)dx\\
 =&\int_{-\infty}^{+\infty}Q\left(N(0, 1)\le \frac{-\ln\left(\frac{\alpha\bar{F}_{2}}{g_{1}\bar{F}_{1}}e^{\sigma_{2}\sqrt{T}
 x}+\frac{K}{g_{1}\bar{F}_{1}}\right)+\sigma_1\rho\sqrt{T}x}{\sigma_1 \sqrt{T}\sqrt{1-\rho^{2}}}\right)\varphi(x)dx\\
 =&\int_{-\infty}^{+\infty}\Phi\left(\frac{-\ln\left(\frac{\alpha\bar{F}_{2}}{g_{1}\bar{F}_{1}}e^{\sigma_{2}\sqrt{T}x}+\frac{K}{g_{1}\bar{F}_{1}}\right)+\sigma_{1}\rho\sqrt{T}x}{\sigma_{1}\sqrt{T}\sqrt{1-\rho^{2}}}\right)\varphi(x)dx.
\end{split}
\end{equation*}
The other two $C_D^2$ and $C_D^3$ can also be evaluated similarly. The expression (\ref{266S}) is obtained by applying Simpson's rule for the Riemann integral in (\ref{new42}). 

\textbf{Proof of formula (\ref{RRR}):} Note that $Y\overset{d}{=}\rho X+\sqrt{1-\rho^2}Z$ for some independent (from $X$) standard normal $Z$. Therefore we have
\begin{equation}
\Pi_T=e^{-rT}E_{X}\left[E_{Z}\left[\left(F_{1}e^{-\frac{1}{2}\sigma_{1}^{2}T+\sigma_{1}\rho\sqrt{T}X+\sigma_{1}\sqrt{1-\rho^{2}}\sqrt{T}Z}-F_{2}e^{-\frac{1}{2}\sigma_{2}^{2}T+\sigma_{2}\sqrt{T}X}-K\right)^{+}|X\right]\right]
\end{equation}
after conditioning with respect to $X$. Observe that
\begin{equation}
\begin{split}
 &E_{Z}\left[\left(F_{1}e^{-\frac{1}{2}\sigma_{1}^{2}T+\sigma_{1}\rho\sqrt{T}x+\sigma_{1}\sqrt{1-\rho^{2}}\sqrt{T}Z}-F_{2}e^{-\frac{1}{2}\sigma_{2}^{2}T+\sigma_{2}\sqrt{T}x}-K\right)^{+}\right]\\
 =&E_{Z}\left[\left(F_{1}e^{-\frac{1}{2}\sigma_{1}^{2}\rho^{2}T+\sigma_{1}\sqrt{T}\rho x}e^{-\frac{1}{2}\sigma_{1}^{2}(1-\rho^{2})T+\sigma_{1}\sqrt{1-\rho^{2}}\sqrt{T}Z}-\left(F_{2}e^{-\frac{1}{2}\sigma_{2}^{2}T+\sigma_{2}\sqrt{T}x}+K\right)\right)^{+}\right]\\
 =&E_{Z}\left[\left(\tilde{F}e^{-\frac{1}{2}\tilde{\sigma}^{2}T+\tilde{\sigma}\sqrt{T}Z}-\tilde{K}\right)^{+}\right]
 \end{split}
\end{equation}
with
\begin{equation}
\begin{split}
 \tilde{F}&=F_{1}e^{-\frac{1}{2}\sigma_{1}^{2}\rho^{2}T+\sigma_{1}\sqrt{T}\rho x},\\
 \tilde{K}&=F_{2}e^{-\frac{1}{2}\sigma_{2}^{2}T+\sigma_{2}\sqrt{T}x}+K,\\
 \tilde{\sigma}&=\sigma_{1}\sqrt{1-\rho^{2}}.
\end{split}
\end{equation}
Note that
\begin{equation}
e^{-rT}E_{Z}\left[\left(\tilde{F}e^{-\frac{1}{2}\tilde{\sigma}^{2}T+\tilde{\sigma}\sqrt{T}Z}-\tilde{K}\right)^{+}\right]\\
 =e^{-rT}\tilde{F}\Phi(d_{1})-e^{-rT}\tilde{K}\Phi(d_{2}),
\end{equation}
where
\begin{equation}
\begin{split}
 \tilde{d}_{1}&=\frac{1}{\tilde{\sigma}\sqrt{T}}\left(\ln\left(\frac{\tilde{F}}{\tilde{K}}\right)+\frac{\tilde{\sigma}^{2}T}{2}\right),\\
 \tilde{d}_{2}&=\frac{1}{\tilde{\sigma}\sqrt{T}}\left(\ln\left(\frac{\tilde{F}}{\tilde{K}}\right)-\frac{\tilde{\sigma}^{2}T}{2}\right).
\end{split}
\end{equation}
Therefore we have
\begin{equation}
\begin{split}
 \Pi_T=&e^{-rT}E\left[\left(F_{1}e^{-\frac{1}{2}\sigma_{1}^{2}T+\sigma_{1}\sqrt{T}Y}-F_{2}e^{-\frac{1}{2}\sigma_{2}^{2}T+\sigma_{2}\sqrt{T}X}-K\right)^{+}\right]\\
 =&e^{-rT}E[\tilde{F}(X)\Phi(\tilde{d}_{1}(X))]-e^{-rT}E[\tilde{K}(X)\Phi(\tilde{d}_{2}(X))].
\end{split}
\end{equation}

\textbf{Proof of the statement of Remark \ref{rem4.44}:}
We have the following relation
\begin{equation}\label{980}
\begin{split}
 &e^{-rT}E[\tilde{F}(X)\Phi(\tilde{d}_{1}(X))]-e^{-rT}E[\tilde{K}(X)\Phi(\tilde{d}_{2}(X))]\\
 =&e^{-rT}F_{1}E\left[e^{-\frac{1}{2}\sigma_{1}^{2}\rho^{2}T+\sigma_{1}\sqrt{T}\rho X}\Phi\left(\frac{1}{\tilde{\sigma}\sqrt{T}}\left(\ln\left(\frac{F_{1}e^{-\frac{1}{2}\sigma_{1}^{2}\rho^{2}T+\sigma_{1}\sqrt{T}\rho X}}{F_{2}e^{-\frac{1}{2}\sigma_{2}^{2}T+\sigma_{2}\sqrt{T}X}+K}\right)+\frac{\tilde{\sigma}^{2}T}{2}\right)\right)\right]\\
 -&e^{-rT}F_{2}E\left[e^{-\frac{1}{2}\sigma_{2}^{2}T+\sigma_{2}\sqrt{T}X}\Phi\left(\frac{1}{\tilde{\sigma}\sqrt{T}}\left(\ln\left(\frac{F_{1}e^{-\frac{1}{2}\sigma_{1}^{2}\rho^{2}T+\sigma_{1}\sqrt{T}\rho X}}{F_{2}e^{-\frac{1}{2}\sigma_{2}^{2}T+\sigma_{2}\sqrt{T}X}+K}\right)-\frac{\tilde{\sigma}^{2}T}{2}\right)\right)\right]\\
 -&e^{-rT}KE\left[\Phi\left(\frac{1}{\tilde{\sigma}\sqrt{T}}\left(\ln\left(\frac{F_{1}e^{-\frac{1}{2}\sigma_{1}^{2}\rho^{2}T+\sigma_{1}\sqrt{T}\rho X}}{F_{2}e^{-\frac{1}{2}\sigma_{2}^{2}T+\sigma_{2}\sqrt{T}X}+K}\right)-\frac{\tilde{\sigma}^{2}T}{2}\right)\right)\right].
\end{split}
\end{equation}
The first term in (\ref{980}) is
\begin{equation}
\begin{split}
 &E\left[e^{-rT}F_{1}e^{-\frac{1}{2}\sigma_{1}^{2}\rho^{2}T+\sigma_{1}\sqrt{T}\rho X}\Phi\left(\frac{1}{\tilde{\sigma}\sqrt{T}}\left(\ln\left(\frac{F_{1}e^{-\frac{1}{2}\sigma_{1}^{2}\rho^{2}T+\sigma_{1}\sqrt{T}\rho X}}{F_{2}e^{-\frac{1}{2}\sigma_{2}^{2}T+\sigma_{2}\sqrt{T}X}+K}\right)+\frac{\tilde{\sigma}^{2}T}{2}\right)\right)\right]\\
 =&e^{-rT}F_{1}E\left[\Phi\left(\frac{1}{\tilde{\sigma}\sqrt{T}}\left(\ln\left(\frac{F_{1}e^{\frac{1}{2}\sigma_{1}^{2}\rho^{2}T+\sigma_{1}\sqrt{T}\rho X}}{F_{2}e^{-\frac{1}{2}\sigma_{2}^{2}T+\rho\sigma_{1}\sigma_{2}T+\sigma_{2}\sqrt{T}X}+K}\right)+\frac{\tilde{\sigma}^{2}T}{2}\right)\right)\right]\\
 =&e^{-rT}F_{1}E\left[\Phi\left(\frac{1}{\tilde{\sigma}\sqrt{T}}\ln\left(\frac{F_{1}e^{\frac{1}{2}\sigma_{1}^{2}T+\sigma_{1}\sqrt{T}\rho X}}{F_{2}e^{-\frac{1}{2}\sigma_{2}^{2}T+\rho\sigma_{1}\sigma_{2}T+\sigma_{2}\sqrt{T}X}+K}\right)\right)\right]\\
 =&e^{-rT}F_{1}P\left(Z\leq\frac{1}{\tilde{\sigma}\sqrt{T}}\ln\left(\frac{F_{1}e^{\frac{1}{2}\sigma_{1}^{2}T+\sigma_{1}\sqrt{T}\rho X}}{F_{2}e^{-\frac{1}{2}\sigma_{2}^{2}T+\rho\sigma_{1}\sigma_{2}T+\sigma_{2}\sqrt{T}X}+K}\right)\right)\\
 =&e^{-rT}F_{1}P\left(F_{1}e^{\frac{1}{2}\sigma_{1}^{2}T+\sigma_{1}\sqrt{T}(\rho X+\sqrt{1-\rho^{2}}Z)}-F_{2}e^{-\frac{1}{2}\sigma_{2}^{2}T+\rho\sigma_{1}\sigma_{2}T+\sigma_{2}\sqrt{T}X}-K\geq0\right)\\
 =&e^{-rT}F_{1}P\left(F_{1}e^{\frac{1}{2}\sigma_{1}^{2}T+\sigma_{1}\sqrt{T}Y}-F_{2}e^{-\frac{1}{2}\sigma_{2}^{2}T+\rho\sigma_{1}\sigma_{2}T+\sigma_{2}\sqrt{T}X}-K\geq0\right),
\end{split}
\end{equation}
where $Z$ represents a standard normal random variable independent from $X$. Similarly, for the second and third terms of (\ref{980}) we have
\begin{equation}
\begin{split}
 &E\left[e^{-rT}F_{2}e^{-\frac{1}{2}\sigma_{2}^{2}T+\sigma_{2}\sqrt{T}X}\Phi\left(\frac{1}{\tilde{\sigma}\sqrt{T}}\left(\ln\left(\frac{F_{1}e^{-\frac{1}{2}\sigma_{1}^{2}\rho^{2}T+\sigma_{1}\sqrt{T}\rho X}}{F_{2}e^{-\frac{1}{2}\sigma_{2}^{2}T+\sigma_{2}\sqrt{T}X}+K}\right)-\frac{\tilde{\sigma}^{2}T}{2}\right)\right)\right]\\
 =&e^{-rT}F_{2}E\left[\Phi\left(\frac{1}{\tilde{\sigma}\sqrt{T}}\left(\ln\left(\frac{F_{1}e^{-\frac{1}{2}\sigma_{1}^{2}\rho^{2}T+\rho\sigma_{1}\sigma_{2}T+\sigma_{1}\sqrt{T}\rho X}}{F_{2}e^{\frac{1}{2}\sigma_{2}^{2}T+\sigma_{2}\sqrt{T}X}+K}\right)-\frac{\tilde{\sigma}^{2}T}{2}\right)\right)\right]\\
 =&e^{-rT}F_{2}E\left[\Phi\left(\frac{1}{\tilde{\sigma}\sqrt{T}}\ln\left(\frac{F_{1}e^{-\frac{1}{2}\sigma_{1}^{2}T+\rho\sigma_{1}\sigma_{2}T+\sigma_{1}\sqrt{T}\rho X}}{F_{2}e^{\frac{1}{2}\sigma_{2}^{2}T+\sigma_{2}\sqrt{T}X}+K}\right)\right)\right]\\
 =&e^{-rT}F_{2}P\left(Z\leq\frac{1}{\tilde{\sigma}\sqrt{T}}\ln\left(\frac{F_{1}e^{-\frac{1}{2}\sigma_{1}^{2}T+\rho\sigma_{1}\sigma_{2}T+\sigma_{1}\sqrt{T}\rho X}}{F_{2}e^{\frac{1}{2}\sigma_{2}^{2}T+\sigma_{2}\sqrt{T}X}+K}\right)\right)\\
 =&e^{-rT}F_{2}P\left(F_{1}e^{-\frac{1}{2}\sigma_{1}^{2}T+\rho\sigma_{1}\sigma_{2}T+\sigma_{1}\sqrt{T}Y}-F_{2}e^{\frac{1}{2}\sigma_{2}^{2}T+\sigma_{2}\sqrt{T}X}-K\geq0\right),
\end{split}
\end{equation}
and
\begin{equation}
\begin{split}
 &E\left[e^{-rT}K\Phi\left(\frac{1}{\tilde{\sigma}\sqrt{T}}\left(\ln\left(\frac{F_{1}e^{-\frac{1}{2}\sigma_{1}^{2}\rho^{2}T+\sigma_{1}\sqrt{T}\rho X}}{F_{2}e^{-\frac{1}{2}\sigma_{2}^{2}T+\sigma_{2}\sqrt{T}X}+K}\right)-\frac{\tilde{\sigma}^{2}T}{2}\right)\right)\right]\\
 =&e^{-rT}KE\left[\Phi\left(\frac{1}{\tilde{\sigma}\sqrt{T}}\ln\left(\frac{F_{1}e^{-\frac{1}{2}\sigma_{1}^{2}T+\sigma_{1}\sqrt{T}\rho X}}{F_{2}e^{-\frac{1}{2}\sigma_{2}^{2}T+\sigma_{2}\sqrt{T}X}+K}\right)\right)\right]\\
 =&e^{-rT}KP\left(Z\leq\frac{1}{\tilde{\sigma}\sqrt{T}}\ln\left(\frac{F_{1}e^{-\frac{1}{2}\sigma_{1}^{2}T+\sigma_{1}\sqrt{T}\rho X}}{F_{2}e^{-\frac{1}{2}\sigma_{2}^{2}T+\sigma_{2}\sqrt{T}X}+K}\right)\right)\\
 =&e^{-rT}KP\left(F_{1}e^{-\frac{1}{2}\sigma_{1}^{2}T+\sigma_{1}\sqrt{T}Y}-F_{2}e^{-\frac{1}{2}\sigma_{2}^{2}T+\sigma_{2}\sqrt{T}X}-K\geq0\right).
\end{split}
\end{equation}

\textbf{Proof of Lemma \ref{lem1}:}
Since $X$ and $Y$ are two standard normal random variables with $Cov(Y,X)=\rho$, we can write $Y$ as $Y=\rho X+\sqrt{1-\rho^{2}}\tilde{\epsilon}$ with $\tilde{\epsilon}\sim N(0,1)$ and $Cov(\tilde{\epsilon},X)=0$. We have 
\begin{equation*}
\begin{split}
 mY-nX&=(m\rho-n)X+m\sqrt{1-\rho^{2}}\tilde{\epsilon}\\
 &\sim N(0,m^{2}+n^{2}-2\rho mn).
\end{split}
\end{equation*}
therefore
\begin{equation*}
\begin{split}
 Q(mY-nX\geq l)&=Q\left(\frac{mY-nX}{\sqrt{m^{2}+n^{2}-2\rho mn}}\geq\frac{l}{\sqrt{m^{2}+n^{2}-2\rho mn}}\right)\\
 &=\Phi\left(\frac{-l}{\sqrt{m^{2}+n^{2}-2\rho mn}}\right).
\end{split}
\end{equation*}\\ \\

\textbf{Proof of Lemma \ref{lem2}:}
We can rewrite the three curves $\mathcal{C}_1, \mathcal{C}_2, \mathcal{C}_3$ as
\begin{equation*}
\begin{split}
 \mathcal{C}_1: \;& y=\frac{1}{\sigma_{1}\sqrt{T}}\ln\left(\frac{\alpha\bar{F}_{2}}{g_{1}\bar{F}_{1}}e^{\sigma_{2}\sqrt{T}x}+\frac{K}{g_{1}\bar{F}_{1}}\right),\\
 \mathcal{C}_2:\; & y=\frac{1}{\sigma_{1}\sqrt{T}}\ln\left(\frac{g_{2}\bar{F}_{2}}{\alpha\bar{F}_{1}}e^{\sigma_{2}\sqrt{T}x}+\frac{K}{\alpha\bar{F}_{1}}\right),\\
 \mathcal{C}_3:\; & y=\frac{1}{\sigma_{1}\sqrt{T}}\ln\left(\frac{\bar{F}_{2}}{\bar{F}_{1}}e^{\sigma_{2}\sqrt{T}x}+\frac{K}{\bar{F}_{1}}\right).
\end{split}
\end{equation*}
For curve $\mathcal{C}_1$,
\begin{equation*}
\begin{split}
 \lim_{x\rightarrow +\infty}\frac{y}{x}&=\frac{1}{\sigma_{1}\sqrt{T}}\lim_{x\rightarrow +\infty}\frac{\ln\left(\frac{\alpha\bar{F}_{2}}{g_{1}\bar{F}_{1}}e^{\sigma_{2}\sqrt{T}x}+\frac{K}{g_{1}\bar{F}_{1}}\right)}{x}\\
 &=\frac{1}{\sigma_{1}\sqrt{T}}\lim_{x\rightarrow +\infty}\frac{\frac{\alpha\bar{F}_{2}}{g_{1}\bar{F}_{1}}e^{\sigma_{2}\sqrt{T}x}\sigma_{2}\sqrt{T}}{\frac{\alpha\bar{F}_{2}}{g_{1}\bar{F}_{1}}e^{\sigma_{2}\sqrt{T}x}+\frac{K}{g_{1}\bar{F}_{1}}}=\frac{\sigma_{2}}{\sigma_{1}},\\
 \lim_{x\rightarrow -\infty}y&=\frac{1}{\sigma_{1}\sqrt{T}}\ln\left(\frac{K}{g_{1}\bar{F}_{1}}\right)=\frac{1}{\sigma_{1}\sqrt{T}}\ln\left(\frac{K}{\bar{F}_{1}}\right)-\sigma_{1}\sqrt{T}.
\end{split}
\end{equation*}
For curve $\mathcal{C}_2$,
\begin{equation*}
\begin{split}
 \lim_{x\rightarrow +\infty}\frac{y}{x}&=\frac{1}{\sigma_{1}\sqrt{T}}\lim_{x\rightarrow +\infty}\frac{\ln\left(\frac{g_{2}\bar{F}_{2}}{\alpha\bar{F}_{1}}e^{\sigma_{2}\sqrt{T}x}+\frac{K}{\alpha\bar{F}_{1}}\right)}{x}\\
 &=\frac{1}{\sigma_{1}\sqrt{T}}\lim_{x\rightarrow +\infty}\frac{\frac{g_{2}\bar{F}_{2}}{\alpha\bar{F}_{1}}e^{\sigma_{2}\sqrt{T}x}\sigma_{2}\sqrt{T}}{\frac{g_{2}\bar{F}_{2}}{\alpha\bar{F}_{1}}e^{\sigma_{2}\sqrt{T}x}+\frac{K}{g_{1}\bar{F}_{1}}}=\frac{\sigma_{2}}{\sigma_{1}},\\
 \lim_{x\rightarrow -\infty}y&=\frac{1}{\sigma_{1}\sqrt{T}}\ln\left(\frac{K}{\alpha\bar{F}_{1}}\right)=\frac{1}{\sigma_{1}\sqrt{T}}\ln\left(\frac{K}{\bar{F}_{1}}\right)-\rho\sigma_{2}\sqrt{T}.
\end{split}
\end{equation*}
For curve $\mathcal{C}_3$,
\begin{equation}\label{74}
\begin{split}
 \lim_{x\rightarrow +\infty}\frac{y}{x}&=\frac{1}{\sigma_{1}\sqrt{T}}\lim_{x\rightarrow +\infty}\frac{\ln\left(\frac{\bar{F}_{2}}{\bar{F}_{1}}e^{\sigma_{2}\sqrt{T}x}+\frac{K}{\bar{F}_{1}}\right)}{x}\\
 &=\frac{1}{\sigma_{1}\sqrt{T}}\lim_{x\rightarrow +\infty}\frac{\frac{\bar{F}_{2}}{\bar{F}_{1}}e^{\sigma_{2}\sqrt{T}x}\sigma_{2}\sqrt{T}}{\frac{\bar{F}_{2}}{\bar{F}_{1}}e^{\sigma_{2}\sqrt{T}x}+\frac{K}{\bar{F}_{1}}}=\frac{\sigma_{2}}{\sigma_{1}},\\
 \lim_{x\rightarrow -\infty}y&=\frac{1}{\sigma_{1}\sqrt{T}}\ln\left(\frac{K}{\bar{F}_{1}}\right).
\end{split}
\end{equation}\\ \\

\textbf{Proof of Proposition \ref{BS-formula}:}
From Proposition \ref{prop1} we have (\ref{C}).
To obtain closed-form formula, we approximate the probabilities in (\ref{C}) by replacing the corresponding curves by lines. Namely we replace $\mathcal{C}_i$ by lines $\ell_i$ for each $i=1,2,3$. For each $i$, we have

\begin{equation*}
\begin{split}
 C_D^i\approx Q\left(y^{i}-b\frac{\sigma_{2}}{\sigma_{1}}x^{i}\geq y_{0}^{i}-b\frac{\sigma_{2}}{\sigma_{1}}x_{0}^{i}\right).
\end{split}
\end{equation*}
Next, we substitute  $y_{0}^{i}$ by $z_{0}^{i}(b\sigma_{2},a,x_{0}^{i})$ for each  $i=1,2,3$ and we use  Lemma \ref{lem1} to  obtain
\begin{equation*}
\begin{split}
 &Q\left(y^{i}-b\frac{\sigma_{2}}{\sigma_{1}}x^{i}\geq y_{0}^{i}-b\frac{\sigma_{2}}{\sigma_{1}}x_{0}^{i}\right)\\
 \approx&Q\left(y^{i}-b\frac{\sigma_{2}}{\sigma_{1}}x^{i}\geq z_{0}^{i}(b\sigma_{2},a,x_{0}^{i})-b\frac{\sigma_{2}}{\sigma_{1}}x_{0}^{i}\right)\\
 =&Q\left(\hat{\epsilon}_{i}\geq\frac{z_{0}^{i}(b\sigma_{2},a,x_{0}^{i})-b\frac{\sigma_{2}}{\sigma_{1}}x_{0}^{i}}{\sqrt{1+\left(b\frac{\sigma_{2}}{\sigma_{1}}\right)^{2}-2\rho b\frac{\sigma_{2}}{\sigma_{1}}}}\right)\\
 =&\Phi\left(\frac{-\delta_{i}}{1+\kappa^{2}-2\rho\kappa}\right).
\end{split}
\end{equation*}\\ \\

\textbf{Proof of Proposition \ref{mainmain}:}
From (\ref{bbb}) and (\ref{asss}), we know that $a_{i}(x)$, $i=1,2,3$, are not constant and they are not necessarily equal to each other, the same is also true for $b_{i}(x)$, $i=1,2,3$. Therefore, by substituting $a$ and $b$ with $a_{1}(\lambda)$, $a_{2}(\mu)$, $a_{3}(\gamma)$ and $b_{1}(\lambda)$, $b_{2}(\mu)$, $b_{3}(\gamma)$ respectively in Proposition \ref{BS-formula} and by following the steps in Proposition \ref{BS-formula}, we  get the result.\\ \\

\textbf{Proof of Corollary \ref{cor52}:}
When $\lambda=(\frac{1}{2}\sigma_{2}-\rho\sigma_{1})\sqrt{T}$,
\begin{equation}
\begin{split}
 a_{1}(\lambda)&=a_{1}\left((\frac{1}{2}\sigma_{2}-\rho\sigma_{1})\sqrt{T}\right)=F_{2}+Ke^{-\sigma_{2}\sqrt{T}(\frac{1}{2}\sigma_{2}-\rho\sigma_{1})\sqrt{T}-\rho\sigma_{1}\sigma_{2}T+\frac{1}{2}\sigma_{2}^{2}T}=F_{2}+K=a,\\
 b_{1}(\lambda)&=b_{1}\left((\frac{1}{2}\sigma_{2}-\rho\sigma_{1})\sqrt{T}\right)=\frac{e^{\rho\sigma_{1}\sigma_{2}T}F_{2}e^{-\frac{1}{2}\sigma_{2}^{2}T}e^{\sigma_{2}\sqrt{T}(\frac{1}{2}\sigma_{2}-\rho\sigma_{1})\sqrt{T}}}
 {e^{\rho\sigma_{1}\sigma_{2}T}F_{2}e^{-\frac{1}{2}\sigma_{2}^{2}T}e^{\sigma_{2}\sqrt{T}(\frac{1}{2}\sigma_{2}-\rho\sigma_{1})\sqrt{T}}+K}=\frac{F_{2}}{F_{2}+K}=b,\\
 \kappa_{1}(\lambda)&=\kappa_{1}\left((\frac{1}{2}\sigma_{2}-\rho\sigma_{1})\sqrt{T}\right)=\frac{\sigma_{2}}{\sigma_{1}}b,
\end{split}
\end{equation}
and
\begin{equation}
\begin{split}
 &z_{0}^{1}(\sigma_{2}b_{1}(\lambda),a_{1}(\lambda),\lambda)\\
 =&\frac{1}{\sigma_{1}\sqrt{T}}\ln\left(\frac{a}{F_{1}}\right)+\frac{1}{\sigma_{1}\sqrt{T}}
 \left(\rho\sigma_{1}\sigma_{2}bT+\sigma_{2}b\sqrt{T}(\frac{1}{2}\sigma_{2}-\rho\sigma_{1})\sqrt{T}-\frac{1}{2}\sigma_{2}^{2}b^{2}T\right)-\frac{1}{2}\sigma_{1}\sqrt{T}\\
 =&\frac{1}{\sigma_{1}\sqrt{T}}\left(\ln\left(\frac{a}{F_{1}}\right)+\frac{1}{2}\sigma_{2}^{2}bT-\frac{1}{2}\sigma_{2}^{2}b^{2}T-\frac{1}{2}\sigma_{1}^{2}T\right).
\end{split}
\end{equation}
Hence
\begin{equation}
\begin{split}
 &\frac{-\delta_1(\lambda)}{\sqrt{1+\kappa_1^2(\lambda)-2\rho\kappa_1(\lambda)}}\\
 =&\frac{-z_{0}^{1}(\sigma_{2}b_{1}(\lambda),a_{1}(\lambda),\lambda)+\kappa_{1}(\lambda)\lambda}{\sqrt{1+\kappa_1^2(\lambda)-2\rho\kappa_1(\lambda)}}\\
 =&\frac{-\frac{1}{\sigma_{1}\sqrt{T}}\left(\ln\left(\frac{a}{F_{1}}\right)+\frac{1}{2}\sigma_{2}^{2}bT-\frac{1}{2}\sigma_{2}^{2}b^{2}T-\frac{1}{2}\sigma_{1}^{2}T\right)
 +\frac{1}{\sigma_{1}\sqrt{T}}\left(\frac{1}{2}\sigma_{2}^{2}bT-\rho\sigma_{1}\sigma_{2}bT\right)}{\sqrt{1+\left(\frac{\sigma_{2}}{\sigma_{1}}b\right)^{2}-2\rho\frac{\sigma_{2}}{\sigma_{1}}b}}\\
 =&\frac{\ln\left(\frac{F_{1}}{a}\right)+\frac{1}{2}\sigma_{1}^{2}T+\frac{1}{2}\sigma_{2}^{2}b^{2}T-\rho\sigma_{1}\sigma_{2}bT}{\sqrt{T}\sqrt{\sigma_{1}^{2}+b^{2}\sigma_{2}^{2}-2\rho\sigma_{1}\sigma_{2}b}}.
\end{split}
\end{equation}
When $\mu=-\frac{1}{2}\sigma_{2}\sqrt{T}$,
\begin{equation}
\begin{split}
 a_{2}(\mu)&=a_{2}\left(-\frac{1}{2}\sigma_{2}\sqrt{T}\right)=F_{2}+Ke^{-\sigma_{2}\sqrt{T}(-\frac{1}{2}\sigma_{2}\sqrt{T})-\frac{1}{2}\sigma_{2}^{2}T}=F_{2}+K=a,\\
 b_{2}(\mu)&=b_{2}\left(-\frac{1}{2}\sigma_{2}\sqrt{T}\right)=\frac{e^{\sigma_{2}^{2}T}F_{2}e^{-\frac{1}{2}\sigma_{2}^{2}T}e^{\sigma_{2}\sqrt{T}(-\frac{1}{2}\sigma_{2}\sqrt{T})}}
 {e^{\sigma_{2}^{2}T}F_{2}e^{-\frac{1}{2}\sigma_{2}^{2}T}e^{\sigma_{2}\sqrt{T}(-\frac{1}{2}\sigma_{2}\sqrt{T})}+K}=\frac{F_{2}}{F_{2}+K}=b,\\
 \kappa_{2}(\mu)&=\kappa_{2}\left(-\frac{1}{2}\sigma_{2}\sqrt{T}\right)=\frac{\sigma_{2}}{\sigma_{1}}b,
\end{split}
\end{equation}
and
\begin{equation}
\begin{split}
 &z_{0}^{2}(\sigma_{2}b_{2}(\mu),a_{2}(\mu),\mu)\\
 =&\frac{1}{\sigma_{1}\sqrt{T}}\ln\left(\frac{a}{F_{1}}\right)+\frac{1}{\sigma_{1}\sqrt{T}}
 \left(\sigma_{2}^{2}bT-\frac{1}{2}\sigma_{2}^{2}bT-\frac{1}{2}\sigma_{2}^{2}b^{2}T\right)+\left(\frac{1}{2}\sigma_{1}-\rho\sigma_{2}\right)\sqrt{T}\\
 =&\frac{1}{\sigma_{1}\sqrt{T}}\left(\ln\left(\frac{a}{F_{1}}\right)+\frac{1}{2}\sigma_{2}^{2}bT-\frac{1}{2}\sigma_{2}^{2}b^{2}T+\frac{1}{2}\sigma_{1}^{2}T-\rho\sigma_{1}\sigma_{2}T\right).
\end{split}
\end{equation}
Hence
\begin{equation}
\begin{split}
 &\frac{-\delta_2(\mu)}{\sqrt{1+\kappa_2^2(\mu)-2\rho\kappa_2(\mu)}}\\
 =&\frac{-z_{0}^{2}(\sigma_{2}b_{2}(\mu),a_{2}(\mu),\mu)+\kappa_{2}(\mu)\mu}{\sqrt{1+\kappa_2^2(\mu)-2\rho\kappa_2(\mu)}}\\
 =&\frac{-\frac{1}{\sigma_{1}\sqrt{T}}\left(\ln\left(\frac{a}{F_{1}}\right)+\frac{1}{2}\sigma_{2}^{2}bT-\frac{1}{2}\sigma_{2}^{2}b^{2}T+\frac{1}{2}\sigma_{1}^{2}T-\rho\sigma_{1}\sigma_{2}T\right)
 +\frac{1}{\sigma_{1}\sqrt{T}}\left(-\frac{1}{2}\sigma_{2}^{2}bT\right)}{\sqrt{1+\left(\frac{\sigma_{2}}{\sigma_{1}}b\right)^{2}-2\rho\frac{\sigma_{2}}{\sigma_{1}}b}}\\
 =&\frac{\ln\left(\frac{F_{1}}{a}\right)-\frac{1}{2}\sigma_{1}^{2}T-\sigma_{2}^{2}bT+\frac{1}{2}\sigma_{2}^{2}b^{2}T+\rho\sigma_{1}\sigma_{2}T}{\sqrt{T}\sqrt{\sigma_{1}^{2}+b^{2}\sigma_{2}^{2}-2\rho\sigma_{1}\sigma_{2}b}}.
\end{split}
\end{equation}
When $\gamma=\frac{1}{2}\sigma_{2}\sqrt{T}$,
\begin{equation}
\begin{split}
 a_{3}(\gamma)&=a_{3}\left(\frac{1}{2}\sigma_{2}\sqrt{T}\right)=F_{2}+Ke^{-\sigma_{2}\sqrt{T}\frac{1}{2}\sigma_{2}\sqrt{T}+\frac{1}{2}\sigma_{2}^{2}T}=F_{2}+K=a,\\
 b_{3}(\gamma)&=b_{3}\left(\frac{1}{2}\sigma_{2}\sqrt{T}\right)=\frac{F_{2}e^{-\frac{1}{2}\sigma_{2}^{2}T}e^{\sigma_{2}\sqrt{T}\frac{1}{2}\sigma_{2}\sqrt{T}}}
 {F_{2}e^{-\frac{1}{2}\sigma_{2}^{2}T}e^{\sigma_{2}\sqrt{T}\frac{1}{2}\sigma_{2}\sqrt{T}}+K}=\frac{F_{2}}{F_{2}+K}=b,\\
 \kappa_{3}(\gamma)&=\kappa_{3}\left(\frac{1}{2}\sigma_{2}\sqrt{T}\right)=\frac{\sigma_{2}}{\sigma_{1}}b,
\end{split}
\end{equation} 
and
\begin{equation}
\begin{split}
 &z_{0}^{3}(\sigma_{2}b_{3}(\gamma),a_{3}(\gamma),\gamma)\\
 =&\frac{1}{\sigma_{1}\sqrt{T}}\ln\left(\frac{a}{F_{1}}\right)+\frac{1}{\sigma_{1}\sqrt{T}}
 \left(\frac{1}{2}\sigma_{2}^{2}bT-\frac{1}{2}\sigma_{2}^{2}b^{2}T\right)+\frac{\sigma_{1}\sqrt{T}}{2}\\
 =&\frac{1}{\sigma_{1}\sqrt{T}}\left(\ln\left(\frac{a}{F_{1}}\right)+\frac{1}{2}\sigma_{2}^{2}bT-\frac{1}{2}\sigma_{2}^{2}b^{2}T+\frac{1}{2}\sigma_{1}^{2}T\right).
\end{split}
\end{equation}
Hence
\begin{equation}
\begin{split}
 &\frac{-\delta_3(\gamma)}{\sqrt{1+\kappa_3^2(\gamma)-2\rho\kappa_3(\gamma)}}\\
 =&\frac{-z_{0}^{3}(\sigma_{2}b_{3}(\gamma),a_{3}(\gamma),\gamma)+\kappa_{3}(\gamma)\gamma}{\sqrt{1+\kappa_3^2(\gamma)-2\rho\kappa_3(\gamma)}}\\
 =&\frac{-\frac{1}{\sigma_{1}\sqrt{T}}\left(\ln\left(\frac{a}{F_{1}}\right)+\frac{1}{2}\sigma_{2}^{2}bT-\frac{1}{2}\sigma_{2}^{2}b^{2}T+\frac{1}{2}\sigma_{1}^{2}T\right)
 +\frac{1}{\sigma_{1}\sqrt{T}}\left(\frac{1}{2}\sigma_{2}^{2}bT\right)}{\sqrt{1+\left(\frac{\sigma_{2}}{\sigma_{1}}b\right)^{2}-2\rho\frac{\sigma_{2}}{\sigma_{1}}b}}\\
 =&\frac{\ln\left(\frac{F_{1}}{a}\right)-\frac{1}{2}\sigma_{1}^{2}T+\frac{1}{2}\sigma_{2}^{2}b^{2}T}{\sqrt{T}\sqrt{\sigma_{1}^{2}+b^{2}\sigma_{2}^{2}-2\rho\sigma_{1}\sigma_{2}b}}.
\end{split}
\end{equation}
This completes the proof.

\bibliographystyle{siam}
\bibliography{Reference}
\end{document}